\documentclass[%
 reprint,
 amsmath,amssymb,
 aps,
]{revtex4-2}

\usepackage{graphicx}% Include figure files
\usepackage{dcolumn}% Align table columns on decimal point
\usepackage{bm}
\usepackage{ulem,color}% bold math
\newcommand{\cE}{{\cal E}}
\newcommand{\apjl}{ApJL}
\newcommand{\Eprint}{Eprint}
\newcommand{\mnras}{MNRAS}

\begin{document}

\title{Gravitational Radiation from Accelerating Jets}% Force line breaks with \\

\author{Elly Leiderschneider}
\author{Tsvi Piran}%
 \email{tsvi.piran@mail.huji.ac.il}
\affiliation{%
Racah Institute for Physics, The Hebrew University, Jerusalem, 91904, ISRAEL  %\textbackslash\textbackslash
}%

\date{\today}% It is always \today, today,

\begin{abstract}
Non-spherical rapid acceleration of mass (or energy) to a relativistic velocity is a natural source of gravitational radiation. Such conditions arise  in both long and short gamma-ray bursts whose central engine ejects relativistic jets. The resulting gravitational wave signal is of a memory type, rising to a finite level (of order $4 G \cE/r$)  over a duration that corresponds to the longer of either the injection time and the acceleration time of the jet. We explore the properties of such signals and their potential detectability. Unfortunately, the expected signals are below the frequency band of Advanced LIGO-Virgo-Kagra, and above LISA. However, they fall within the range of the planned BBO and DECIGO. While current sensitivity is marginal for the detection of jet gravitational wave signals from GRBs, hidden relativistic jets that exist within some core collapse SNe could be detected. Such a detection would reveal the acceleration mechanism and the activity of the central engine, which cannot be explored directly in any other way. 
\end{abstract}

\maketitle

%\tableofcontents
\section{Introduction}
\label{sec:introduction}

Gamma-ray bursts (GRBs) are extremely energetic, with typical energies of $\cE=10^{51}$ erg. Jets associated with a GRB are accelerated to high Lorentz factors, with ${\Gamma\gtrsim 100}$ being a typical value. They are highly anisotropic, with the ejected material being confined to a cone with opening angle ${\theta_{\rm j}\lesssim 10^o}$.  These jets are accelerated from rest within a short time, and they last for fraction of a second (in short GRBs) to tens of seconds (in long ones). 
The acceleration of a relativistic jet produces a memory-type gravitational wave (GW) signal \cite{segalis,piran2002}. 
{ Observations of this GW signal will reveal  the nature of the jets and  the acceleration process. 
Additionally, there is  ample evidence for  hidden jets activity within some supernovae \cite{Kulkarni98,MacFadyen01,Tan01,Soderberg06,Bromberg11,Soderberg06,Bromberg11,Izzo19,Nakar19}. The  existence of these jets can be inferred only indirectly.  A detection of this kind of  GW signal is possibly the only direct way to identify these invisible jets  and learn about their hidden features. }

While the GW amplitude estimates \cite{piran2002,Sago,ofek}, that are of order $10^{-24}-10^{-25}$ for reasonably nearby GRBs, 
and the relevant frequencies (that are in the decihertz range) both make detection difficult, it is worthwhile to get back to this problem, and explore in greater details both the characteristics and the detection prospects of the GW signal.

{ Segalis and Ori \cite{segalis} and Piran \cite{piran2002}
considered an instantaneously accelerated point particle, using  the zero-frequency limit (ZFL). This approximation that corresponds to infinite acceleration  is appropriate for describing the final jump in the amplitude of the GW. However, this approximation misses, naturally, the details of the temporal structure that are crucial for consideration of detection feasibility. }

{ Sago et al. \cite{Sago} generalized this result for a GRB model based on a large number of thin jets (``minijets") \cite{Yamazaki2004} that are ejected at random angles within a cone and random times within the duration of the GRB. Within this model each minijet produces a single pulse and these pulses combine to form the GRB light curve. In their model each minijet is described by an instantaneously accelerated  point particle generating a step function signal. 
The superposition of the different step functions results in a complicated GW light curve. The model captures the effects of the angular structure and of the overall duration of the GRB resulting  in a typical time scale for the pulse rise time that is comparable to the duration of the burst.  }

{  Birnholtz and Piran \cite{ofek} relaxed the instanteneous acceleration approximation   and developed a scheme for calculating the  GW signal from a continuously accelerating axisymmetric jet. The considered, following the fireball model \cite{Shemi1990}, an acceleration model  in which the jet's Lorentz factor  increases linearly with time (or distance) until it reaches its final value. They considered different angular structures and observers at different viewing angles taking into account integration  over  equal arrival time surfaces. The combined effects of prolonged acceleration  and  taking into account the integration over the arrival time surface results in a temporal structure of the order of the acceleration time at viewing angles close to the jet and longer at larger angles. }

{ In this work, we calculate the GW emission from accelerating jetc combining both effects of prolonged acceleration and prolonged duration of ejection of the jet. } We calculate  properties of the GW that are universal and independent of  particular acceleration models, and combine them with a realistic possible model of the ejection of outflow in GRBs  to derive typical amplitudes and detection expectation of GRBs and other  astrophysical jets. In the following we will be using G=c=1, but at times we introduce these coefficients for clarity.  

{ The structure of the paper is as follows. We outline in \S \ref{sec:Instanteneou} the general description of the problem and following the methods  of \cite{ofek} (that consider instantaneous injection) and some results of \cite{Sago} (that consider instantaneous acceleration) we describe the GW signals from systems with instantaneous ejection or instantaneous acceleration. We explore in \S \ref{sec:temporal}  the temporal structure focusing on the interplay between the two time scales  that exist in the system, the acceleration time scale, $t_{\rm acc}$,  and the overall duration of the activity of the central engine that accelerates the jet, $t_{\rm inj}$. We consider in \S \ref{sec:GRBs} an example in use the temporal structure of GRBs' light curves as a proxy for the activity of the central engine. Following this example we consider in \S \ref{sec:detectability} the detectability of these signals and we summarize and discuss our results in \ref{sec:discussion}.}

%\section{Description of the problem} 
%\label{sec:description}
\section{Instantaneous Ejection and Acceleration}
\label{sec:Instanteneou}

\begin{figure}
\includegraphics[width=0.25\textwidth]{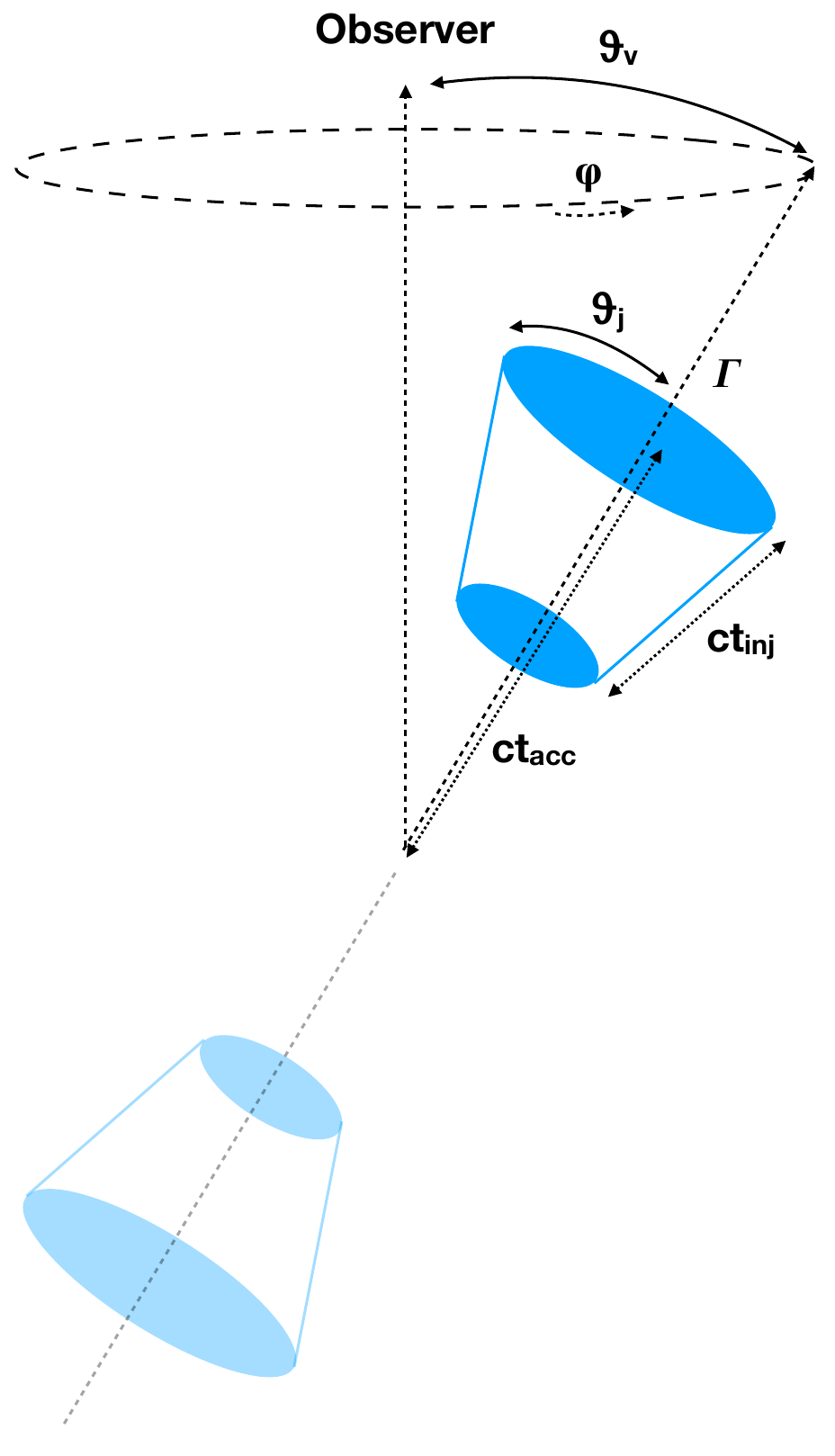}
\caption{A schematic description of the jet. The top shell has reached the final Lorentz factor at a distance $c t_{\rm acc}$ from the origin. The duration of mass injection is $t_{\rm inj}$. A counter-jet is shown in light colors. }
\label{fig:schematic}
\end{figure}

We consider  an idealized  jet that is  accelerated to an ultra-relativistic velocity.  
The jet has  energy  ${\cal E}= m \Gamma $, with  $m$ the jet's mass and $\Gamma$ its  final Lorentz factor. To simplify the discussion we keep only the essential features of the problem (see Fig. \ref{fig:schematic}). The  jet is an axisymmetric top hat with an opening angle $\theta_{\rm j}$. 
The jet moves radially outwards, and every particle emitted at the same time accelerates in the same manner. Particles  emitted at the same time maintain the shape of a radially expanding infinitesimally thin spherical cap. 
The observer is located at a distance $r$  and at an angle $\theta_{\rm v}$, relative to the jet's symmetry axis.

The energy (or mass)  ejection function,  $\dot m(t)$,  describes  the rate  of mass ejection, where $t$ is measured in the rest frame of the central engine, and is the same in the observer's rest frame.
The function $\dot m(t)$ is characterized by the timescale $t_{\rm inj} $.
The acceleration is described by the  function $\Gamma(t)$, where  $t$ is measured in the central engine's frame of reference. 
$\Gamma(t)$ is characterized by the  acceleration timescale  $t_{\rm acc}$.
The time of flight scale that characterizes the arrival time from different angular regions of the jet is related to the acceleration time as 
\begin{equation}
    \tilde t_{\rm o}(\theta_{\rm v}) = (1-\beta \cos\Delta \theta_{\rm v}) t_{\rm acc} ,
    \label{eq:t_obs}
\end{equation}
where $\Delta \theta_{\rm v}$ is the ``relevant'' (as discussed later)  angle between the observer and the source and $\beta$ is the jet's velocity.
As the critical time scale is the longer of the two we denote $t_{\rm c} \equiv {\rm max}(\tilde t_{\rm o},t_{\rm inj} )$. As  $\tilde t_{\rm o}$ depends on the viewing angle, the dominant time scale may be $t_{\rm inj} $ for some observers and $t_{\rm acc}$ for others.

Among the different approximations, the zero-frequency limit (ZFL)  stands out \cite{segalis,piran2002}. This approximation ignores the detailed temporal structure of the source and the corresponding GW signal.
The acceleration and mass ejection are instantaneous: $t_{\rm acc}=0$, and $t_{\rm inj} =0$.  While non-physical, this limit gives an idea of the emerging patterns. It is also relevant for low-frequency detectors whose response is slower than the relevant timescales of the system. 
The waveform, in this limit, is described by a Heaviside step function:
\begin{equation}
\label{eq:theta}
    h(t, \theta_{\rm v}) = h_0(\theta_{\rm v}) \mathcal H (t) \ 
\end{equation}  
 and its Fourier transform is given by 
\begin{equation}
\tilde h(f , \theta_{\rm v}) ={h_0(\theta_{\rm v})}/{f} \ .
\label{eq.FT}
\end{equation} 
%A gravitational wave detector operating at frequencies that are $\ll t_{\rm acc}^{-1},t_{\rm inj} ^{-1}$ will measure a discontinuous jump in the GW amplitude. 

\subsection{A Point Particle - $\theta_{\rm j}=0$ and $t_{\rm inj} =0$}

 We begin considering a point particle of mass $m$ that is instantaneously accelerated to a Lorentz factor $\Gamma$ so that the total energy is ${\cal E} = m \Gamma$. The particle is  moving at polar angles $\theta_{\rm v}$ and $\phi$ in the observer's frame of reference (see Fig. \ref{fig:schematic}). 
%\begin{equation}
%    \vec \beta =  \beta \begin{pmatrix}  \cos \theta_{\rm v} \cos\phi \\ \cos \theta_{\rm v} \sin \phi \\ \sin \theta_{\rm v} \end{pmatrix} ,
%\end{equation}
The gravitational wave amplitudes $h_{\rm +}$ and  $h_{\rm x}$ of  the two polarization modes are given by \cite{segalis}:
\begin{equation}
\label{eq:phase}
    h^{TT} (\theta_{\rm v}) = h_{\rm +} + i h_{\rm x} = \frac{2 \cE \beta^2}{r} \frac{\sin^2 \theta_{\rm v}}{1-\beta \cos\theta_{\rm v}} e^{2 i \phi}.
\end{equation}
For a single point-particle, the phase, $2 i\phi$, can be ignored. When discussing the metric perturbation of an ensemble of particles, though, the complex phase may lead to destructive interference, and one component of the perturbation tensor may dominate over the other.
%The maximal amplitude  (of either the '+' or 'x' components) is given by \cite{segalis}:
%\begin{equation}
%    \label{eq:point_particle}
%    h(\theta_{\rm v}) = \frac{2\cE  \beta^2}{r} \frac{\sin^2\theta_{\rm v}}{1-\beta \cos\theta_{\rm v}} \ .
%\end{equation}

The angular dependence  of the amplitude $h(\theta_{\rm v})$ exhibits 
 \textit{anti-beaming}: the GW amplitude vanishes along its direction of motion, and remains small at a cone around it. It reaches 50\% of the maximal values at an opening angle 
%{$\theta_{a-b} = $}
$\Gamma^{-1}$. The function $h(\theta_{\rm v})$ attains a maximum of
\begin{equation}
    h_{\rm max} = \frac{4{\cE} }{r} ,   \ \ \ \ {\rm at } \ \ \  \theta_{\rm max}=\sqrt{{2}/{\Gamma}} \ .
   \label{eq:hmax}
\end{equation}

The total  GW energy emitted   is given by: 
\begin{equation}
    E_{\rm GW} = \frac{1}{32 \pi} \iint {\dot h}^2 dt d\Omega \ ,
\end{equation}
where $\dot \ $ denotes time derivative.
For an instantaneously accelerating particle, this integral diverges. However, this divergence is not physical, and it arises from the  instantaneous approximation. 
For a finite acceleration time $t_{\rm acc}$ or a finite injection time the temporal integral can be calculated in Fourier space:
\begin{equation}
    E_{\rm GW}= \frac{1}{32 \pi} \int d\Omega \int _0 ^{f(\theta_{\rm v})} \tilde h(f)^2 f^2 d f \ , 
    \label{eq:Energy} 
\end{equation}
where $f(\theta_{\rm v}) = \min (t_{\rm  j}^{-1},\tilde t_{\rm o}^{-1})$ (with $\tilde t_{\rm o}$ calculated here using $\Delta \theta_{\rm v}= \theta_{\rm v}$)  is the angle-dependent upper cutoff on the frequency given by the finite acceleration and injection times.  Integrating we obtain \cite{ofek}:
\begin{equation} 
E_{\rm GW} =\cE^2 \begin{cases} \frac{1}{2 t_{\rm acc}} \left[\frac{3-\beta^2}{\beta} \ln \frac{1+\beta}{1-\beta}-6 \right]  &\mbox{if } \tilde t_{\rm o}> t_{\rm inj} \ ,   \\
\frac{2}{t_{\rm inj} } \left[(2 -\frac{4\beta}{3}) + \frac{1-b^2}{\beta} \ln \frac{1+\beta}{1-\beta}\right] & \mbox{if }  t_{\rm inj}   > \tilde t_{\rm o}  \  . \end{cases} 
\label{eq:E_gamma}
\end{equation} 
{ When adding the coefficients $G$ and $c$ this expression becomes 
$E_{\rm GW} \propto [G \cE /c^4 \max(t_{\rm acc},t_{\rm inj})] \cE $.}

The ratio of the GW emitted energy to the total energy of the particle, $\cE$, 
vanishes when $\beta \rightarrow 0$. 
%The factor $2\cE/{t_{\rm acc}} < 1 $ as the acceleration time is limited by $(G/c^3) M \Gamma$. 
However, if $t_{\rm acc}$ is the dominant (longest) time scale
it diverges when $\Gamma \rightarrow \infty$. Namely, the accelerating engine deposits, in such a case,  more energy in generating gravitational radiation than in accelerating the jet.  
If the jet is self-accelerating this is of course impossible, but then the acceleration process has to be considered more carefully\footnote{Note that in this case $t_{\rm acc} \le \cE$, however the term in square brackets can still be larger than unity.}.
%If $t_{\rm acc}$ is the dominant (longest) time scale this ratio can be larger than unity. 

\begin{figure}
\includegraphics[width=0.45\textwidth]{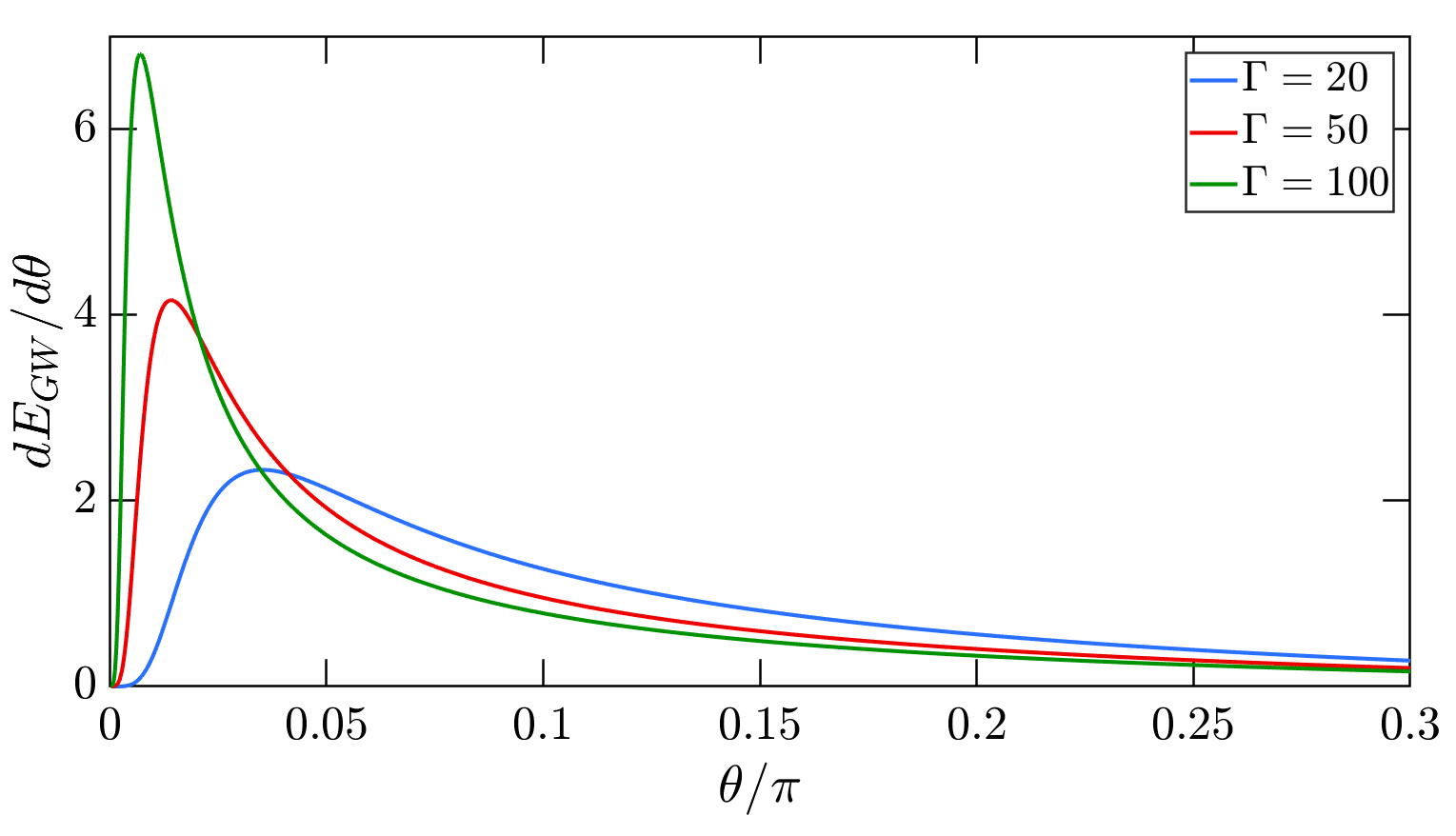}
\caption{The angular distribution of the normalized GW energy for three different Lorentz factors. Energy is beamed in the forward direction, such that 50\% of the GW energy is confined in a cone with opening angle $\sqrt{{2}/{\Gamma}}$. The area under all the distributions is normalized to unity.}
\label{fig:de_dtheta}
\end{figure}

While the GW amplitude, $h$,   is anti-beamed, the GW energy  is  beamed in the forward direction (see Fig. \ref{fig:de_dtheta}).
50\% of the GW energy is deposited in a cone with an opening angle $\theta_{50\%} = \sqrt{{2}/{\Gamma}}$. This may seem counter-intuitive at first. One must remember, however, that  while the GW amplitude decreases over an angular scale $\Gamma^{-1}$ around the axis, the observed frequency of the GW is also boosted in this direction. When both effects are taken into account we find that, while very little energy is emitted within the anti-beamed  cone of $\Gamma^{-1}$,   the overall energy is still beamed in the forward direction just around this inner cone.

Instantaneous ejection of two point particles in opposite directions will lead to a wave form that is the sum of the two 
\begin{equation}
\label{eq:phase}
    h (\theta_{\rm v}) =  \frac{4 \cE \beta^2}{r} \frac{ 1-\cos^2 \theta_{\rm v}}{1-\beta^2 \cos\theta_{\rm v}^2} \ .
\end{equation}
$h (\theta_{\rm v})$ is almost flat apart  from the minima along the two axes. However, the energy  is still beamed in cones of width $\sqrt{2/\Gamma}$, as the contribution of the particle that is moving away from the observer will be seen only at much lower frequencies than the one moving towards it. 

\subsection{A Narrow stream - $\theta_{\rm j}=0$ and $t_{\rm inj}  \ne 0$}

Relaxing somewhat the ZFL approximation, we generalize the previous results to a continuous ejection of a narrow stream over $t_{\rm inj}$. 
All one needs to do is to integrate the single particle $h$   (Eq. \ref{eq:phase}) over the emission time.  As all particles contribute with the same phase, there is no destructive interference. The final jump in the GW signal remains the same and so is the angular structure and the maximal viewing angle. 
There are though two important differences.   The amplitude increases  following $m(t)$ on a time scale $t_{\rm inj}$. 
(see \S \ref{sec:temporal}  below). This results in a typical frequency of $1/ t_{\rm inj}$ that determines both the temporal structure of $h$ and the total energy emitted, as  already discussed in Eq. \ref{eq:E_gamma}. 

\subsection{A Spherical Cap  - $\theta_{\rm j}\ne 0$ and $t_{\rm inj}  = 0$ }

%\del{As mentioned in section \ref{sec:prelim}, we assume that all particles ejected at the same time will undergo the same acceleration process.  Every particle in the jet is also assumed to move on a radial trajectory. Therefore, tracking all the particles ejected at a certain moment, their shape will be that of a radially expanding spherical cap. \\}

We consider next a thin spherical cap  of particles ejected simultaneously and accelerated instantaneously.  
The cap is defined by its opening angle $\theta_{\rm j}$, final Lorentz factor $\Gamma$, total energy $\cE$, and the angle between its center and  the observer $\theta_{\rm v}$. We will  assume that the cap is wide, namely $\Gamma^{-1} \ll \theta_{\rm j}$. Otherwise if   $\theta_{\rm j} \lesssim \Gamma^{-1}$ the signal converges to the point-particle limit. 

We define the observer's line of sight to the emitting source as the $z$ axis of our coordinate system. The coordinates $\theta$ and $\phi$ are defined in the observer's coordinate system in the usual manner (see Fig. \ref{fig:schematic}). Without loss of generality, we define the direction of the jet as $(\theta,\phi)=(\theta_{\rm v} , 0)$ in the observer's frame of reference.  

%\del{If the jet's density is constant over a single shell at a given time, its GW signal should therefore be symmetric under the transformation $\phi \rightarrow -\phi$. }
%\note{I thought that the fact that the jet is axisymmetric }

\begin{figure}
\includegraphics[width=0.25\textwidth]{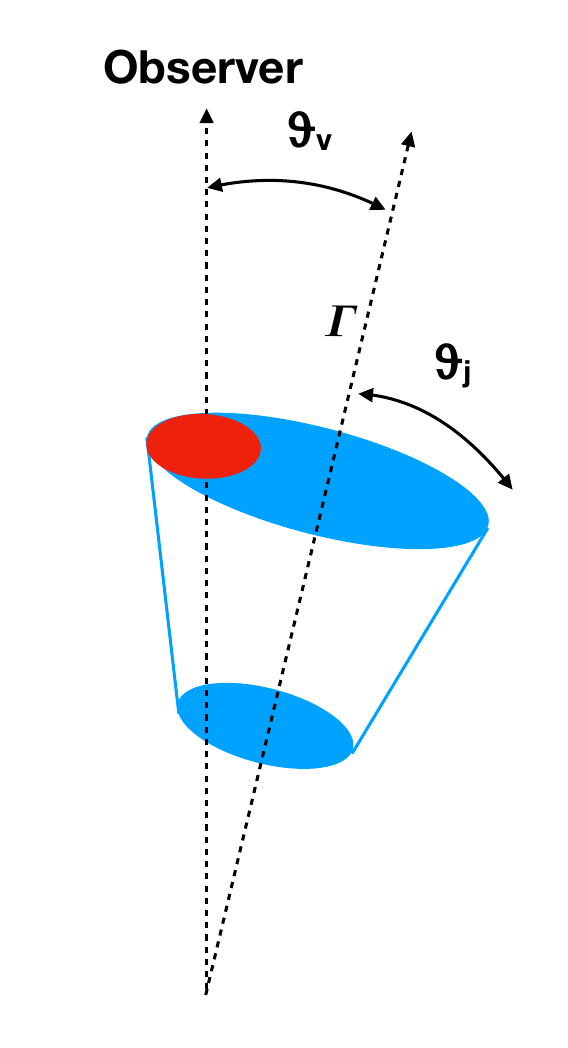}
\caption{A schematic view  of the jet (blue) for $\theta_{\rm v} <\theta_{\rm j}$. Due to the symmetry, the contribution to the GW amplitude of the  part of the jet that is spherically symmetric around the observer
(shown in red ) vanishes. The amplitude from partial rings with $\theta>\theta_{\rm j}-\theta_{\rm v}$,   is reduced compared to the amplitude of a point-particle with the same energy and angle to the observer.
The jet is  symmetric under the transformation $\phi \rightarrow -\phi$: hence, the metric perturbation component $h_{\rm x}$ vanishes identically.}
\label{fig:rings}
\end{figure}

The axial symmetry implies a symmetry under the transformation $\phi \rightarrow -\phi$. Therefore, the metric perturbation $h_{\rm x}$ (which is now summed over the shell) vanishes identically (see Eq. \ref{eq:phase}, and Fig. \ref{fig:rings}). In the following, we simply denote $h=h_{\rm +}$, the only non-vanishing component of the metric perturbation tensor.

Integrating over the cap we find:
\begin{equation}
\label{eq:h_int}
    h_{\rm cap}(\theta_{\rm v} , \theta_{\rm j}) =\frac{2\cE \beta^2}{r \Delta \Omega} \int_{|\theta_{\rm v}-\theta_{\rm j}|} ^{min(\theta_{\rm j}+\theta_{\rm v},\pi)}   \frac{\sin^3 \theta \cdot \sin 2\Delta \phi}{1-\beta \cos\theta}  d\theta \ , 
\end{equation}
where $\Delta \Omega \equiv 2 \pi (1-\cos\theta_{\rm j}^2)$, the solid angle of the cap, and 
\begin{equation}
\label{eq:dphi}
    \Delta \phi \equiv  \cos^{-1} \left[ \frac{\cos \theta_{\rm j}-\cos \theta_{\rm v} \cos\theta}{\sin\theta_{\rm v} \sin\theta} \right]  \ .
\end{equation}

\begin{figure}
\includegraphics[width=0.45 \textwidth]{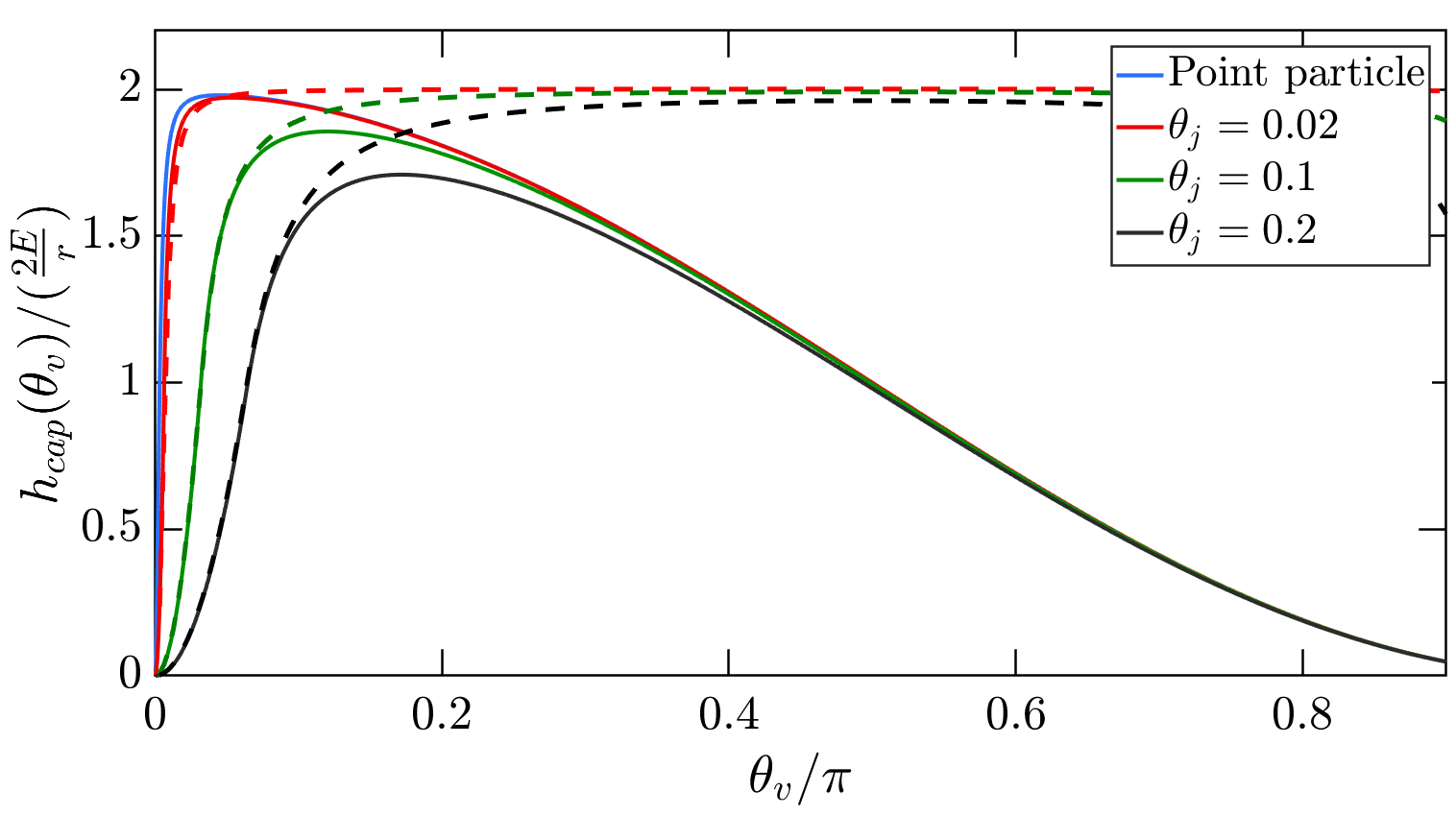}
\includegraphics[width=0.45\textwidth]{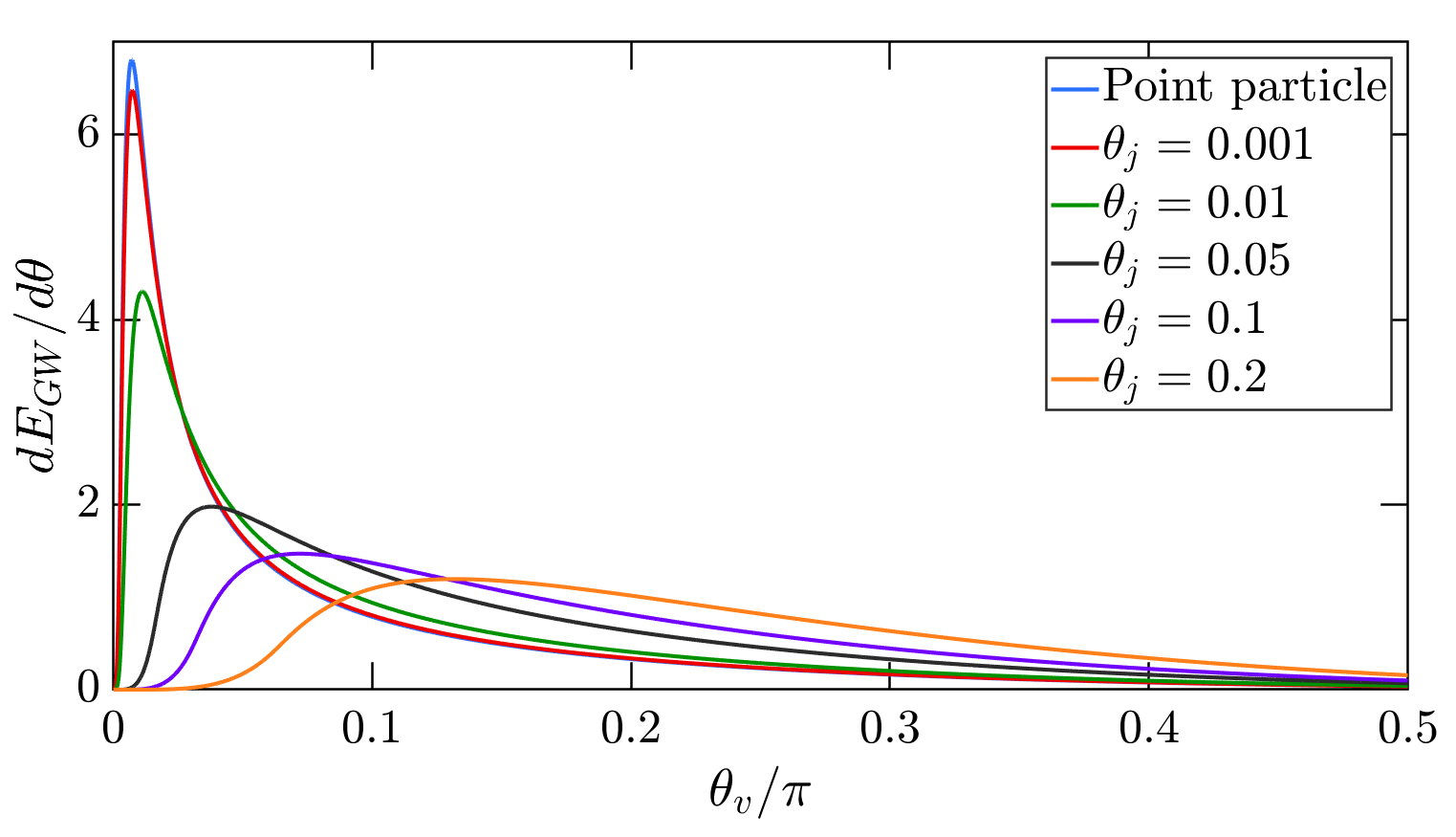}
\caption{The angular distribution of $h$,  the GW amplitude (top) and $dE_{\rm GW}/d\theta$, the normalized energy distribution (bottom) from an accelerating spherical cap with $\Gamma=100$. The anti-beaming region is  $ \approx 0.84 \ \theta_{\rm j}$.  Note the different angular scale of the two figures. The area under each energy distribution is normalized to unity. Dashed lines in the top figure represent the amplitudes of double-sided jets.}
\label{theta_j_vs_pp}
\end{figure}

\begin{figure}
\includegraphics[width=0.45\textwidth]{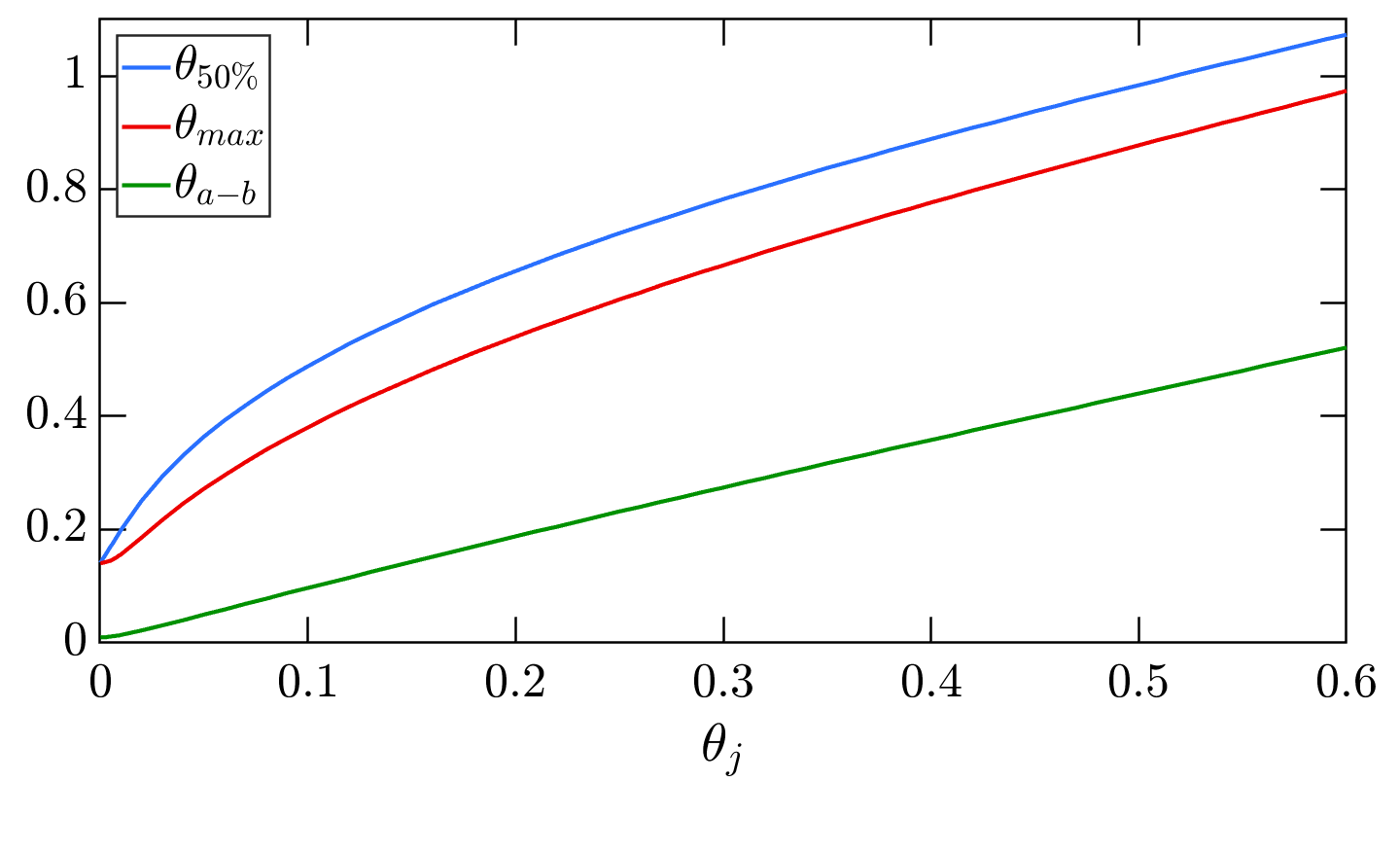}
\caption{Specific viewing  angles for a jet with $\Gamma=100$ as a function of $\theta_{\rm j}$: $\theta_{50\%}$ is the opening angle of the cone which constrains 50\% of the GW's energy, $\theta_{\rm max}$ is the viewing angle with the maximal observed GW amplitude, and $\theta_{a-b}$ is the anti-beaming angle, at which the GW amplitude drops to 50\% of maximum. All plots are with $\Gamma=100$. For $\Gamma^{-1} \ll \theta_{\rm j}$, all three angles are determined by $\theta_{\rm j}$. The intercepts with the $\theta_{\rm j}=0$ axis are determined by the point-particle results.}
\label{fig:thetamax_and_ab}
\end{figure}
\begin{figure}
\includegraphics[width=0.45\textwidth]{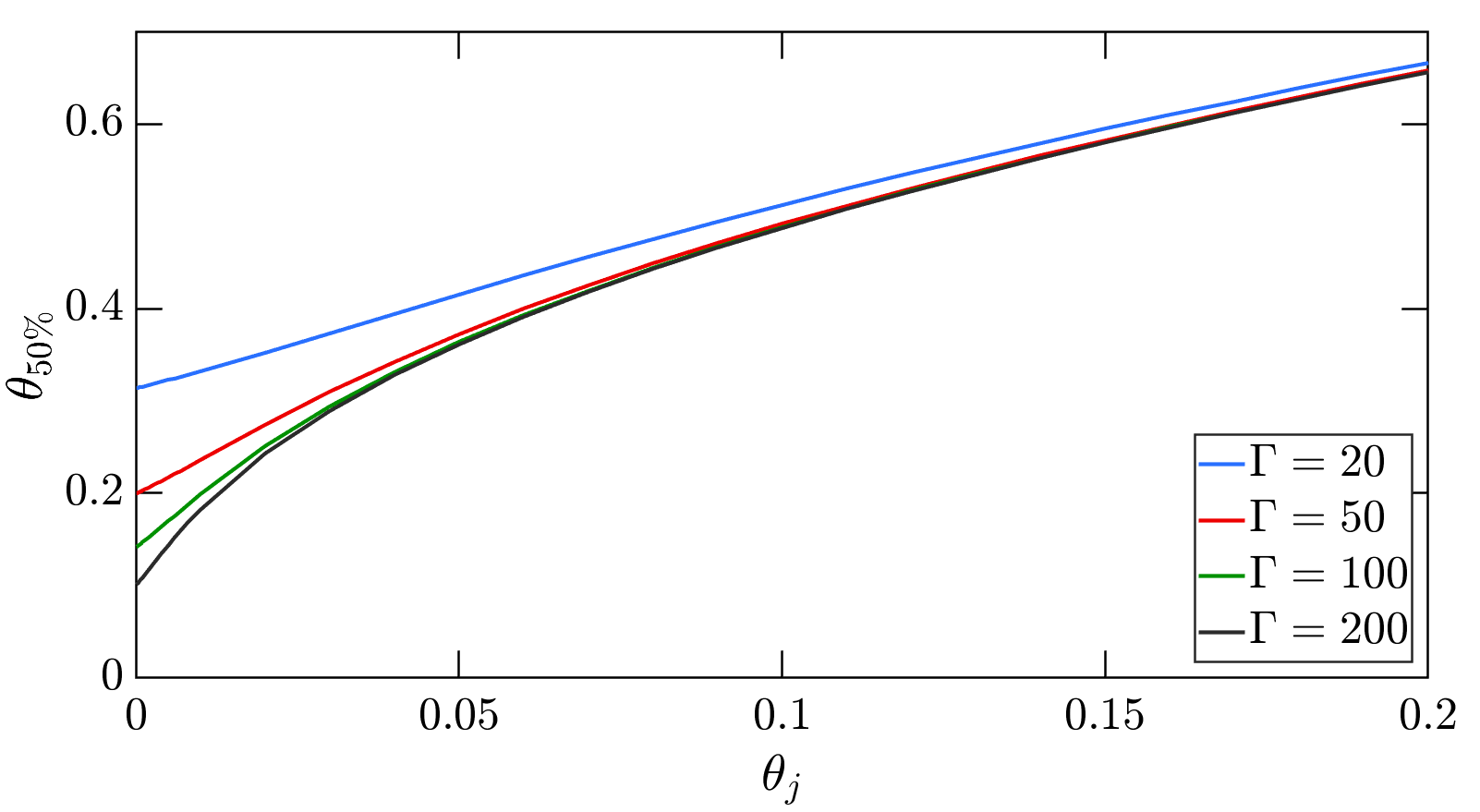}
\caption{The GW energy beaming angle, $\theta_{50 \%}$, as a function of the jet's opening angle, for jets with different Lorentz factors. The intercepts with the $\theta_{\rm j}=0$ axis correspond to $\sqrt{{2}/{\Gamma}}$, but for $\Gamma^{-1} \ll \theta_{\rm j}$, the angle $\theta_{50\%}$ is determined by $\theta_{\rm j}$. Note that the corresponding energy distribution is peaked around $\theta_{\rm j}$ and $\pi-\theta_{\rm j}$}
\label{fig:theta50}
\end{figure}

%\begin{figure}
%\includegraphics[width=0.45\textwidth]{de_dtheta_jet.png}
%\caption{The angular distributions of the GW energy for jets with different opening angles for $\Gamma=100$. The area under each distribution is normalized to unity. }
%\label{fig:edist_thetaj}
%\end{figure}

Figure \ref{theta_j_vs_pp} depicts $h_{\rm cap}(\theta_{\rm v},\theta_{\rm j})$ for different opening angles. This   angular behavior resembles the point-particle result, with a major difference: the anti-beaming region,
which was $ \Gamma^{-1}$ in the point-particle case, is now  $  \approx 0.84 \ \theta_{\rm j}$  (see Fig. \ref{fig:thetamax_and_ab}),
and it is independent of $\Gamma$. This is due to the fact that any region of the cap which is axially symmetric around the observer 
would have no contribution to the GW amplitude. For $\theta_{\rm v} < \theta_{\rm j}$, 
only the outer region of the cap, with $\theta> \theta_{\rm j} - \theta_{\rm v}$,  
contributes. The effect is twofold: regions of the cap with $\theta<\theta_{\rm j}-\theta_{\rm v}$  
have a vanishing contribution to the amplitude, and even in the outer region, destructive interference between symmetric regions will reduce the GW amplitude. 

The maximal GW amplitude is now a function of $\theta_{\rm j}$ (compare with Eq. \ref{eq:hmax}). For small opening angles:
\begin{equation}
    h_{\rm max}(\theta_{\rm j}) \approx \frac{4 \cE}{r} (1-\frac{3}{4} \theta_{\rm j}) \ .
\end{equation}
%The maximal amplitude occurs at an angle
%\begin{equation} 
%\theta_{\rm max} \approx ???
%\end{equation}
%which is determined by $\theta_{\rm j}$ 
%(see Fig. \ref{fig:thetamax_and_ab}). 

Using the amplitude $h_{\rm cap}(\theta_{\rm v} , \theta_{\rm j})$, we calculate the total GW energy, as a straightforward generalization of Eq. \ref{eq:E_gamma}, but now for simplicity we estimate it only for $\tilde t_{\rm o}$ in which we use $\Delta \theta_{\rm v} = \theta_{\rm v}+\theta_{\rm j}$:
\begin{equation}
    E_{\rm cap}(\theta_{\rm j}) =\frac{1}{16 t_{\rm acc}} \int_{0} ^\pi \frac{h_{\rm cap} (\theta_{\rm v},\theta_{\rm j})^2}{1-\beta \cos(\theta_{\rm v}+\theta_{\rm j})} \sin\theta_{\rm v} d\theta_{\rm v}  \ . 
    \label{eq:E_integral_jet}
\end{equation}
{Again the energy will diverge for a strictly instantaneous acceleration. To estimate the energy in a realistic case, we have to introduce a frequency cutoff that depends on the acceleration time. But because of time of flight effects, it also depends on the relation between the viewing angle and the opening angle of the jet and the final velocity:
  $\tilde t_{\rm o}= ({1-\beta \cos(\theta_{\rm v}+\theta_{\rm j})}){t_{\rm acc}}.$ }

Similarly to the case of the GW amplitude anti-beaming angle, we find that the angle of the cone which constrains 50\% of the cap GW's energy, $\theta_{50 \%}$,
 is determined by $\theta_{\rm j}$ and not by $\Gamma$. Fig. \ref{fig:thetamax_and_ab} depicts the GW amplitude's anti-beaming angle, as well as $\theta_{50\%}$ and the angle $\theta_{\rm max}$ where the observed GW amplitude is maximized, all as a function of the jet's opening angle $\theta_{\rm j}$. Fig. \ref{fig:theta50} shows $\theta_{50 \%}$ as a function of  $\Gamma$. For $\Gamma^{-1} \ll \theta_{\rm j}$, the energy beaming angle is determined only by $\theta_{\rm j}$. Fig. \ref{theta_j_vs_pp} shows the angular distribution of the GW energy for jets with different opening angles.

\subsection{Double-sided jets}

The angular distribution of the GW signal changes drastically if the jet is two-sided. We turn to examine two spherical caps of equal energy that are accelerated  along two opposite directions. In this case,  the GW amplitude is a monotonically increasing function of $\theta_{\rm v}$, up to $\pi/2$, where it is maximal  (see Fig. \ref{theta_j_vs_pp}).  The maximal GW amplitude is: 
\begin{equation}
\label{eq:hmax_2head}
    h_{\rm max} (\theta_{\rm j}) = \frac{4 \cE}{r} \cos\theta_{\rm j},
\end{equation}
where $\cE$ is now the total energy of both caps.
%\begin{figure}
%\includegraphics[width=0.45\textwidth]{theta_j_vs_pp_for_2head_jet.png}
%\caption{The angular distribution of a double-headed jet. The maximal amplitude will be measured by an observer with a line of sight perpendicular to the jet's axis of symmetry. All distributions have $\Gamma=100$.}
%\label{2head_distribution}
%\end{figure}
The result is similar to the case of two ejected point particles but, similarly to the single-cap case, the width of the suppressed area around the axes is now of order $ 0.84 \theta_{\rm j}$, rather than of order $\sqrt{2/\Gamma}$.

\section{The Temporal Structure}
\label{sec:temporal}
We turn now to consider the effect of the more detailed temporal structure of the source and the acceleration process. 

\subsection{Power Spectrum and Timescales}
 
%The critical time scale is the longest one. It  is either the acceleration timescale $t_{\rm acc}$ or the ejection timescale $t_{\rm inj} $.

A memory-type signal, rising to an asymptotic value $h_0 (\theta_{\rm v})$ over a  timescale $t_{\rm c}
(=\max[t_{\rm acc},t_{\rm inj})]) $, has a characteristic Fourier transform:
\begin{equation}
    \label{eq:crossover}
    \tilde h(f,\theta_{\rm v}) =  \begin{cases}  {h_0(\theta_{\rm v})}/{f} , \quad f \leq f_{\rm c}  \\ {h_0(\theta_{\rm v}) f_{\rm c} } {g(f)} ,\ \ f \geq f_{\rm c}  \end{cases}
\end{equation}
where $f_{\rm c}  \equiv {1}/{t_{\rm c}}$ is the crossover frequency and $g(f)$, which depends on the nature of the source, decreases faster than  $1/f$.  %$\alpha>1$. 
As the total GW energy must be finite, the  integral $\int_{0} ^\infty {\dot h}^2 (t) dt = \int _0 ^\infty f^2 {\tilde h}^2 (f) df$ yields an asymptotic bound of $g \propto f^{-\alpha_{\inf}}$ with $\alpha_{\inf} >3/2$. 

%At low  frequencies, the Fourier transform of any gravitational wave memory signal will resemble Eq. \ref{eq:theta}. If the detector's frequency is lower than the equivalent frequency for the longest timescale of the burst, one cannot discern any temporal structure in the GW signal. The detector will measure a step function, and the Fourier signal will behave like $f^{-1}$. Around $f_{{\rm c}}$ there is a smooth transition to  $f^{-\alpha}$ behavior at higher frequencies. 

The Fourier transform is closely related to the \textit{spectral density}, which is typically used to characterize the signal-to-noise ratio of the GW:
\begin{equation}
    S(f) \equiv \tilde h(f) \cdot \sqrt{f}
    \label{eq:density}
\end{equation}  
%Eq. \ref{eq:crossover} implies that 
The combination of the crossover frequency, $f_{\rm c}$,  and the spectral density at this frequency,  $S(f_{\rm c}) $, is critical to determine the detectability of the signal. The condition
\begin{equation}
S_{det} (f )< S(f_{\rm c} ) (f_{\rm c} /f)^{1/2}
\label{eq:Sfc}
\end{equation}  
is a necessary but not sufficient condition for the detection of this signal. 
For a low-frequency detector (with a typical frequency range below $f_{\rm c} $) this condition is sufficient as it will detect such event if Eq. \ref{eq:Sfc} is satisfied for some frequency in its range, $f$.
This detector will observe a step function. As $S(f)$ decreases faster than $f^{-1/2}$, Eq. \ref{eq:Sfc} is not a sufficient condition for detection by a high frequency detector. 
If it is sensitive enough, a higher frequency detector  can detect the relevant and interesting temporal structure that exist beyond a simple step function.

As the signal can  be characterized by the crossover frequency, the following sections are concerned with identifying this frequency in the GW's Fourier spectrum.
We  discuss first the simplifying limit $t_{\rm inj}  = 0$, in which the jet is emitted at once. We then examine the general case, of a finite $t_{\rm inj} $. 
%\del{The results of the $t_{\rm inj} =0$ limit will, of course, coincide with the general case whenever $t_{\rm acc} \ll t_{\rm inj} $.}

\subsection{Instantaneously spherical cap - $t_{\rm acc} = 0$, $t_{\rm acc}=0$}

We consider here the GW signal of a single spherical cap, of angular size $\theta_{\rm j}$, that is instantaneously injected and accelerated, but the acceleration takes place at a radius $R$  rather than at the origin.   
We decompose the spherical cap to concentric rings around the observer. The signal from a full ring vanishes. The signal from a partial ring at an angle $\theta $ to the observer is a Heaviside step function, whose magnitude and phase are characterized by $l(\theta) e^{2 i \Delta \phi}$, where  $l(\theta)$ is the fraction of the ring within the cap (see Fig. \ref{fig:rings}) and $\Delta \phi$, defined by Eq. \ref{eq:dphi}, is the corresponding phase.

The arrival time of the signal from this ring is $(1-\beta \cos \theta) R/c$.  Integration over these (partial) rings yields the GW signal:
\begin{align}
     \label{eq:cap_fourier}
     \tilde h_{\rm cap}(f) = \frac{2\cE \beta^2}{r \Delta \Omega} \int_{|\theta_{\rm v}-\theta_{\rm j}|} ^{\theta_{\rm v}+\theta_{\rm j}} d\theta \frac{sin^3 \theta}{1-\beta \cos\theta} l(\theta) e^{2 i \Delta \phi} \nonumber \\
    \times \frac{i}{f} e^{i f (1-\beta \cos\theta) R/c} ,
 \end{align}
where the lower integration limit is determined by the requirement that the GW contribution of a whole ring vanishes, and $ \exp[{i f (1-\beta \cos\theta) R/c}]/{f}$ is the Fourier transform of the Heaviside function.

 The crossover frequency for this GW signal is determined by the time delay between the earliest and latest components of the signal: 
  \begin{equation}
      \label{eq:fc_cap}
      f_{\rm c}  = \frac{1}{\cos(\theta_{\rm v}-\theta_{\rm j})-\cos(\theta_{\rm v}+\theta_{\rm j})} \frac{c}{R} \ . 
  \end{equation}
 Note that generalizing this result to a non-instanteous acceleration we can use $R = c t_{\rm acc} $. 
%\add{This has a simple physical interpretation: the crossover frequency is . In this case, since the cap is emitting instantaneously at a radius $R$, the earliest observed GW signal will be that which is emitted from the point in the cap which is closest to the observer, i.e. $\theta=\theta_{\rm v}-\theta_{\rm j}$. The latest signal will be that emitted from the most distant point in the cap, $\theta=\theta_{\rm v}+\theta_{\rm j}$. 
%The crossover frequency is determined by the delay between these two signals: a measurement at a lower operating frequency will miss out on the temporal details of the signal entirely, and observe only a discontinuous jump.}

\subsection{Continuously accelerating spherical cap -  $t_{\rm acc} \ne 0$. }

The signal from a  cap that is accelerating continuously depends on the specific acceleration model.
Birnholtz \& Piran \cite{ofek} calculated $h(t)$ for a cap accelerating according to the basic fireball GRB model \cite{Goodman86,Piran_fireball}, $\Gamma \propto R$. 

Repeating their calculations for different   $(\theta_{\rm v} , \theta_{\rm j})$, we find (see Fig. \ref{fig:fourier_cap_1} and also Fig. 11 of \cite{ofek}) that  the 
corresponding crossover frequency is given by the time delay between the earliest ($t=0$ at the origin) and latest ($\theta = \theta_{\rm v}+\theta_{\rm j}$ at the end of the acceleration phase) signals:
\begin{equation}
    f_{\rm c ~ |{t_{\rm inj}=0}} = \frac{1}{1-\beta \cos(\theta_{\rm v}+\theta_{\rm j})} \frac{1}{t_{\rm acc}}
    \label{eq:fc_fireball}
\end{equation}

\begin{figure}
\includegraphics[width=0.45\textwidth]{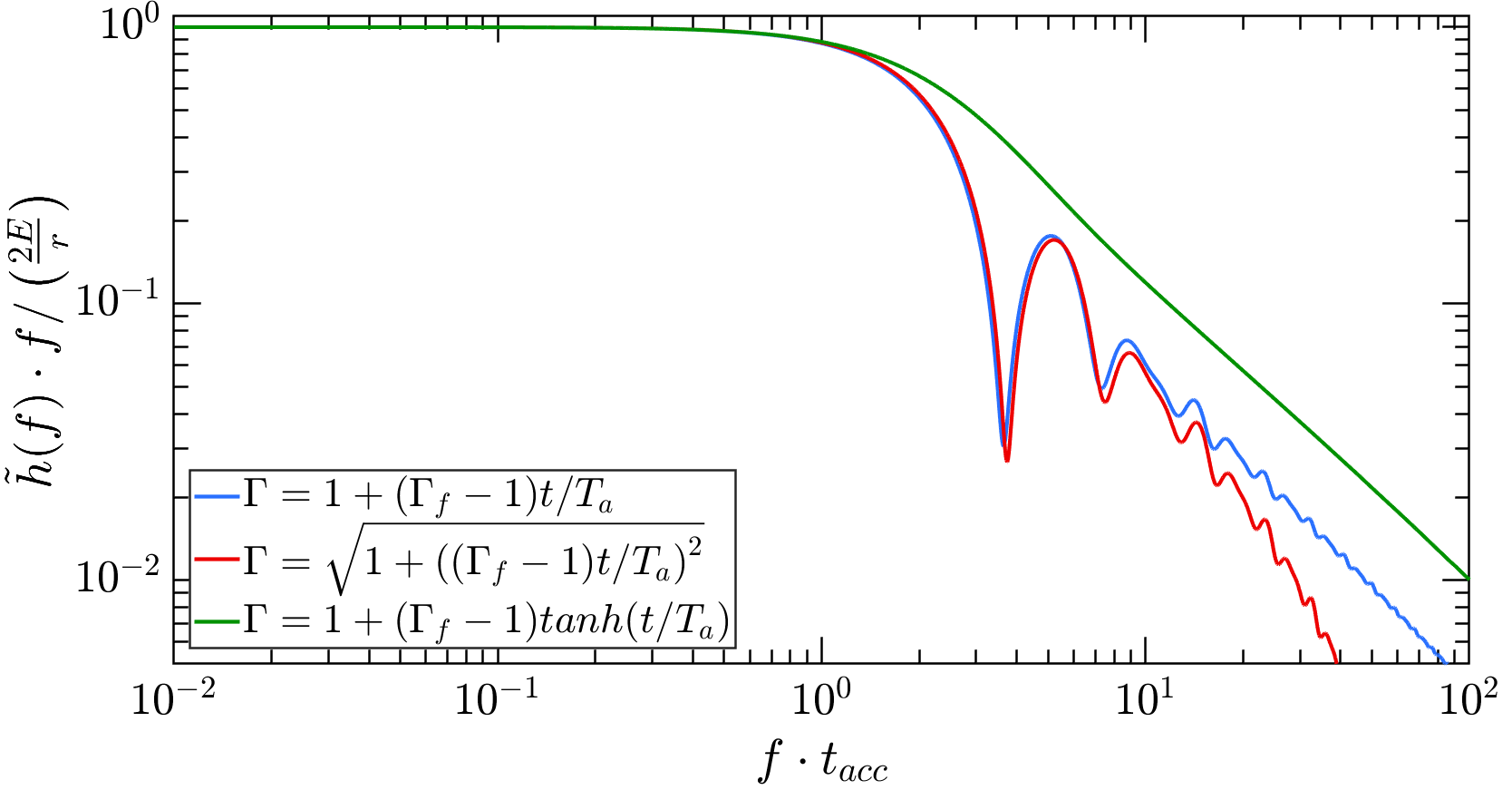}
\caption{The Fourier transforms of $f$ from jets with  three different acceleration models (with  $\Gamma_f = 100$, $\theta_{\rm j} = 0.1$, $\theta_{\rm v} = 0.9$). The time constant of the third model was chosen such that the high-frequency power laws would coincide.}
\label{fig:acc_models}
\end{figure}

\begin{figure}
\includegraphics[width=0.45\textwidth]{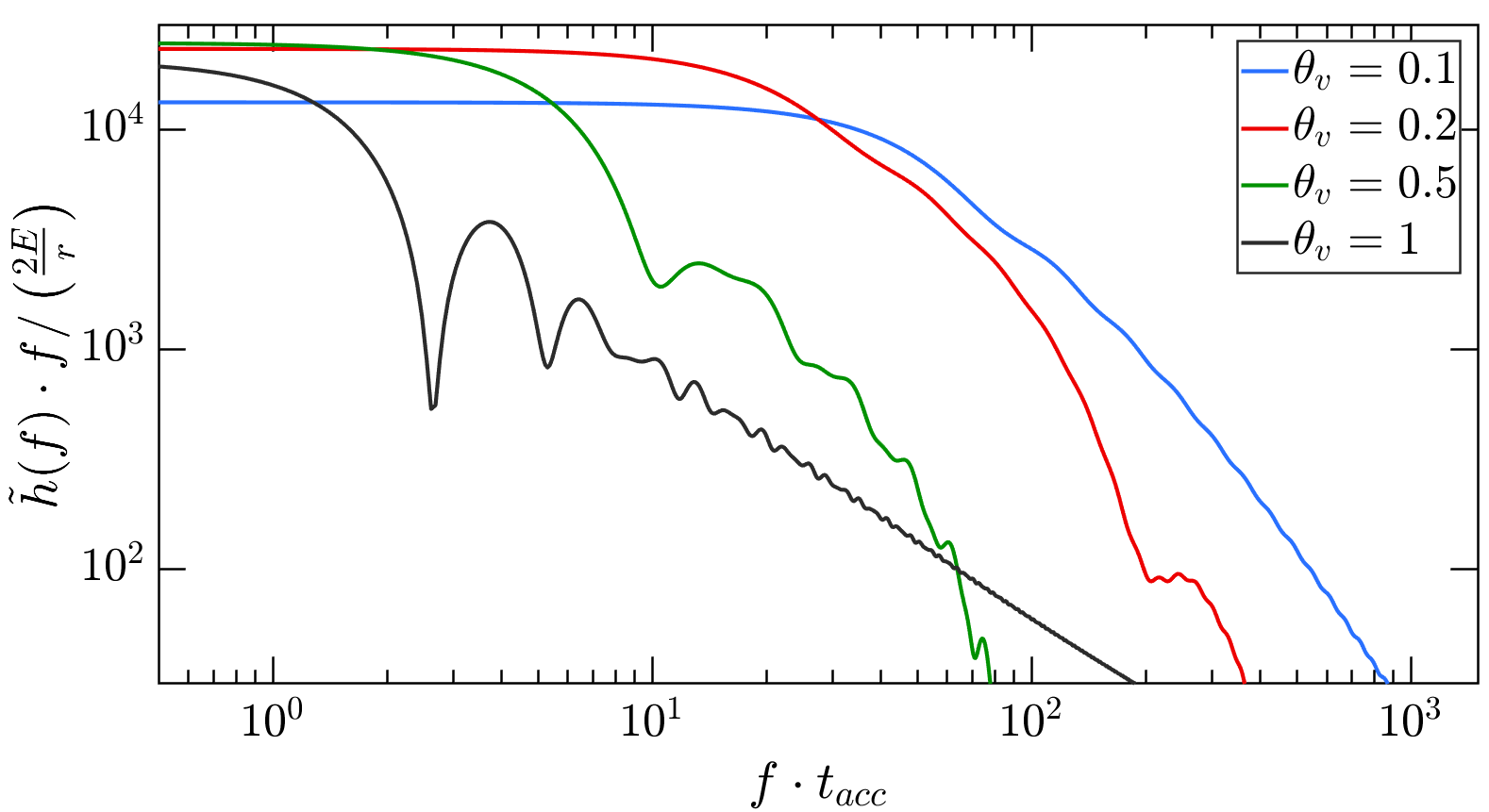}
\caption{The normalized Fourier waveform multiplied by the frequency for jets with the acceleration model $\Gamma \propto R$, based on the numerical code describe in \cite{ofek} for  
$\Gamma=100$ and $\theta_{\rm j}=0.1$. Below the crossover frequency, 
$\tilde h(f) \cdot f$ is a constant. } 
\label{fig:fourier_cap_1}
\end{figure}

While this result was derived for a specific acceleration model, Eq. \ref{eq:fc_fireball} is quite general, being derived purely from geometrical arguments. {We plot in Fig. \ref{fig:acc_models} the Fourier transforms of GWs based on three different acceleration models: the fireball model $\Gamma(t)-1 = (\Gamma-1) t / t_{\rm acc} $; a  constant acceleration in the jet's frame of reference $\Gamma(t) ^2-1 = (\Gamma -1)^2 (t / t_{\rm acc} ) ^2$, and $\Gamma(t)-1  =  (\Gamma-1) \tanh(t / t'_{\rm acc})$. For all three models, we find that the final jump in amplitude is indeed given by the ZFL limit in Eq. \ref{eq:h_int}, and that the crossover frequencies are given by Eq. \ref{eq:fc_fireball}.}

{For all three models considered, we find that the high-frequency behavior, $g(f)$ is described by a power law $f^{-\alpha}$, with $\alpha \approx 2$. For a constant acceleration, the low-frequency behavior coincides with the fireball model. This is no surprise, since the equivalent long-timescale acceleration in both cases is $\Gamma(t) \sim t$. For the third acceleration model, the hyperbolic function's typical timescale is not defined as clearly, so we tuned its timescale parameter, $t'_{\rm acc}$, such that the high-frequency power law would coincide with the two other models.}

\subsection{The crossover diagram}
\label{sec:crossover}

\begin{figure}
\includegraphics[width=0.45\textwidth]{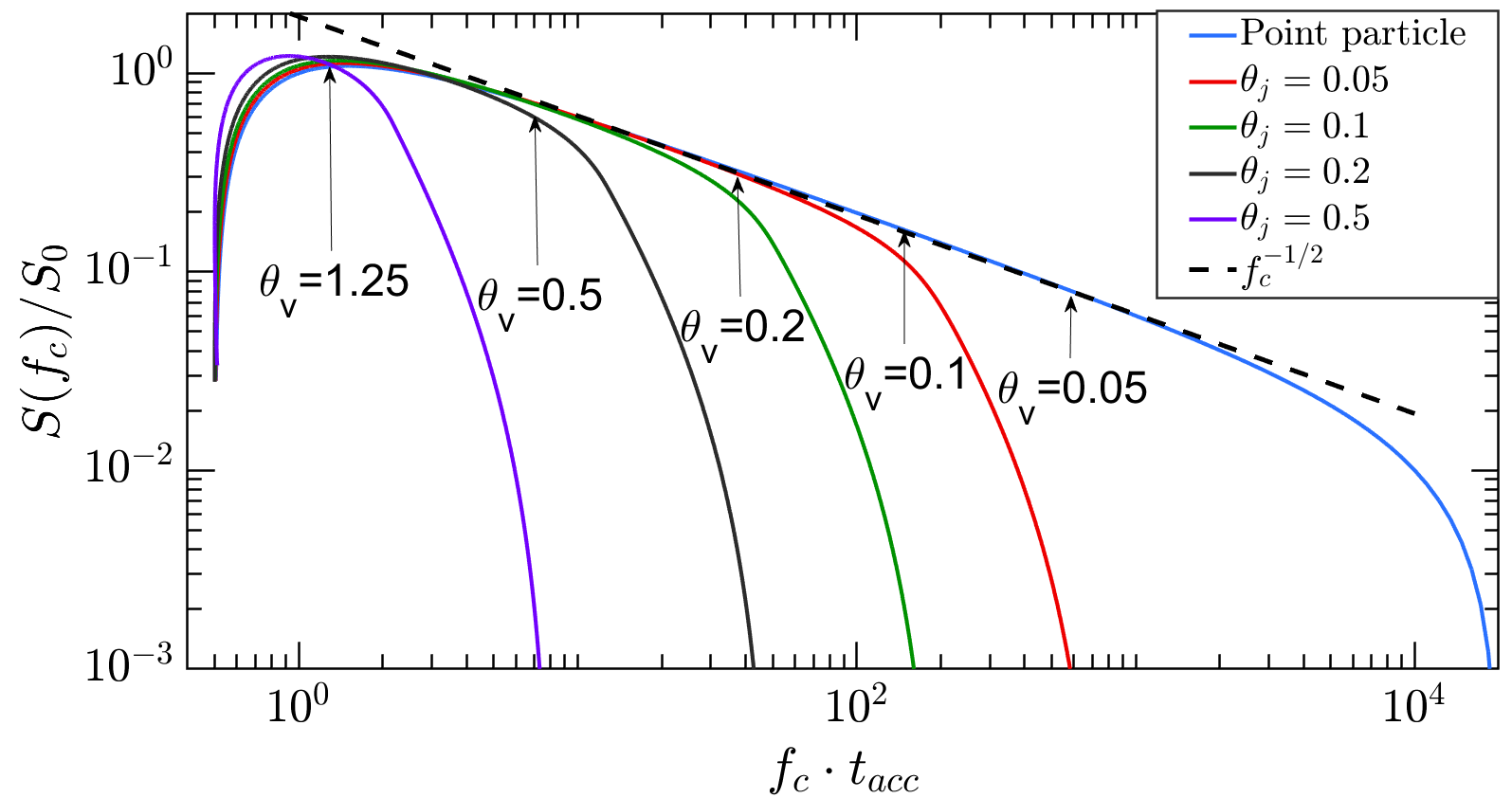}
\caption{The crossover diagrams of jets with different opening angles, compared with that of a point particle for different opening angles (and a point particle) for  $\Gamma=100$ and $t_{\rm acc}=1 {\rm sec}$. The
spectral density is normalized by the value $S_0 \equiv ({2 \cE}/{r}) \sqrt{t_{\rm acc}}$. Along the curves, both the crossover frequency, $f_{\rm c} $, and the spectral density at that frequency, $S(f_{\rm c} )$, vary as a function of the observer angle $\theta_{\rm v}$. Some observer angles are indicated for reference. }
\label{fig:crossover_pp}
\end{figure}

The observed frequency of the GW  can get boosted by a maximal factor of $2\Gamma^2$ along the direction of motion of the jet.  However, because of the {anti-beaming}, the signal is minimal in that direction. 
%Given a detector with an observable frequency band $f_{min} < f < f_{\rm max}$, can this boost the GW signal of a given source %with $t_{\rm acc} ^{-1}< f_{min}$ into the observable frequency range? 
%To address this question, we show in 
Fig.  \ref{fig:crossover_pp} depicts the \textit{crossover diagram}, $S(f_{\rm c} )$ vs. $f_{\rm c} $, for different jets. This diagram represents how the observed spectral density varies as it is viewed from different viewing angles. 
%Figure \ref{fig:crossover_pp} shows the crossover diagrams for single-headed jets with equal energies and different opening angles (along with the limiting case of a point particle, i.e. $\theta_{\rm j}=0$. 
 
Figure \ref{fig:crossover_pp} demonstrates that the jet's finite opening angle reduces the possible boost in the crossover frequency from $ {2\Gamma^2}$ to $ ({1-\cos\theta_{\rm j}})^{-1}$. 
While the boost in frequency increases the jet's crossover frequency, it is always accompanied by a reduction in the observed spectral density, since the angular region in which the frequency is boosted is  well within the GW's anti-beaming region, meaning that the overall spectral density at high frequencies is diminished. 
The  spectral density is comparable to the maximal value over a wide range of viewing angles. For example, for  small opening angles ($\theta_{\rm j} < 0.3$), and with $\Gamma=100$, $S(f)$ is maximal at an observer angle $\theta_{\rm v} \approx \cos^{-1} ({1}/{3}) = 1.23$,
and it exceeds 50\% of the maximum value for $0.38<\theta_{\rm v}<2.17$, corresponding to  75\% of the sky.

The crossover diagram of a double-headed jet, consisting of two jets propagating in two opposite directions, is rather similar to that of a single jet. This is, again, due to the anti-beaming of the GW signal. The jump in amplitude for each jet component is determined by Eq. \ref{eq:h_int}.  For small observer angles, the amplitude of the jet propagating away from the observer will be negligible compared to the amplitude of the jet heading towards the observer (see Fig.  \ref{theta_j_vs_pp}). The two jets will have comparable amplitudes  only in the intermediate range of observer angles, $\theta_{\rm v} \approx {\pi}/{2}$. The contribution of both jets in this angular range is slightly higher than that of a single jet.
%(see Fig. \ref{fig:crossover_double_jet}). 

%\begin{figure}
%\includegraphics[width=0.45\textwidth]{crossover_for_double_jet.png}
%\caption{The crossover diagram of a single jet, compared to that of a double-headed jet with equal energy. $\theta_{\rm j}=0.1,\Gamma=100, t_{\rm acc}=1 sec$ were used for both jets. For a double-headed jet, the range of observer angles is $0 \leq \theta_{\rm v} \leq \pi / 2$, which is why its crossover plot ends abruptly at the point equivalent to $\theta_{\rm v}=\pi / 2$. }
%\label{fig:crossover_double_jet}
%\end{figure}

\subsection{$t_{\rm inj}  \neq 0$}
\begin{figure}
\includegraphics[width=0.45\textwidth]{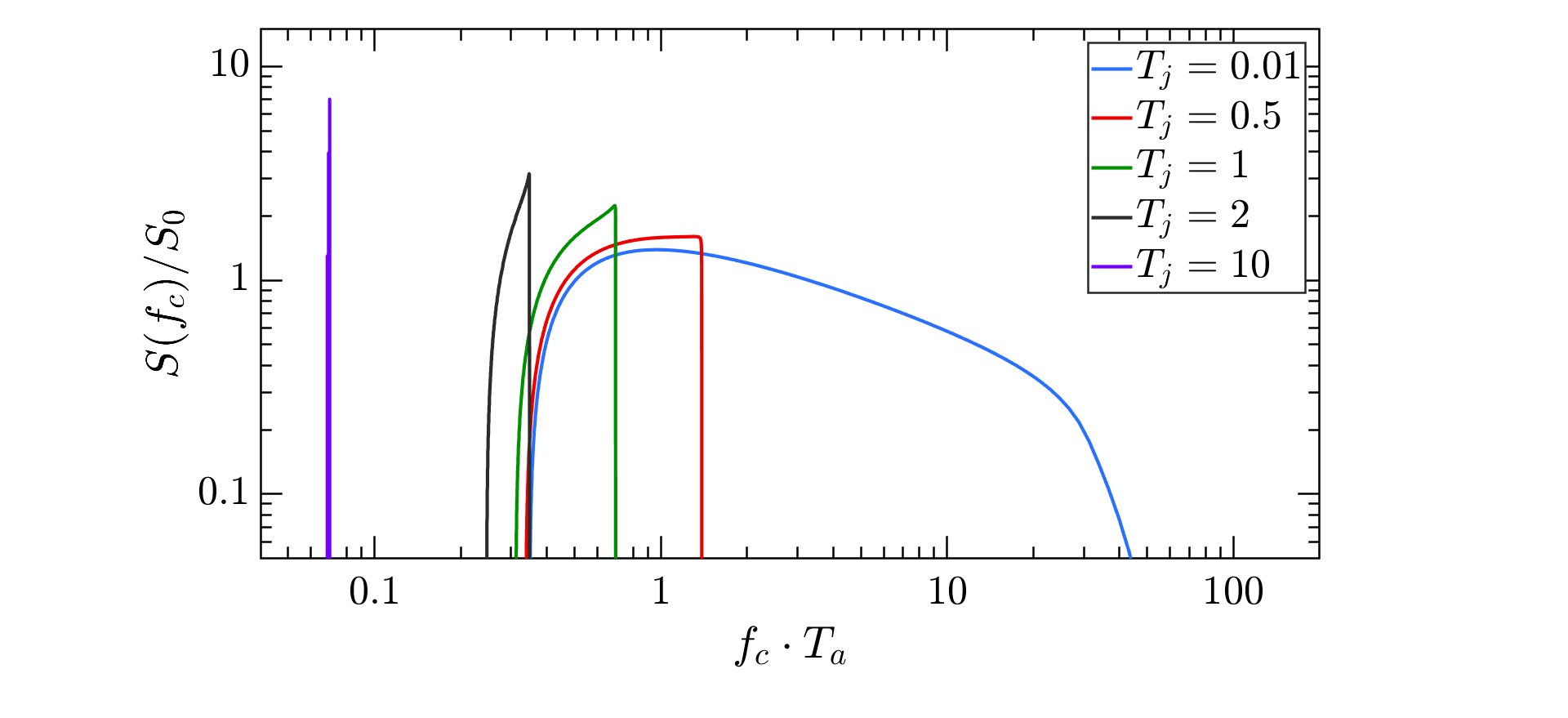}
\caption{The crossover diagrams for jets with both $t_{\rm acc}$ and $t_{\rm inj} $. We keep $t_{\rm acc}$ fixed , with $\Gamma=100$ and $\theta_{\rm j}=0.1$ for all diagrams, and vary $t_{\rm inj} $. The amplitude $h_0(\theta_{\rm v}, \theta_{\rm j})$ is given by Eq. \ref{eq:h_int} and the frequency $f_{\rm c} $ is then extracted from the Fourier transform of Eq. \ref{eq:toy}.  }
\label{fig:tj_and_t0}
\end{figure}

{
With the introduction of another timescale $t_{\rm inj} $, the problem becomes more complex. The main point of the previous section, though, is unchanged: the Fourier transform of the signal is monotonically decreasing, and the crossover from $1/f$ behavior to a steeper decrease occurs at a frequency $f_{\rm c} $. The only difference between this and the $t_{\rm inj} =0$ case is the way in which the crossover frequency $f_{\rm c} $ is determined. The situation is complicated, though, because the timescale determined by the acceleration, $[1- \beta \cos(\theta_{\rm v}+\theta_{\rm j})]t_{\rm acc}$  is angle-dependent. While $t_{\rm inj}$ can be larger for some angles, the acceleration-related timescale can be larger for others. 

To demonstrate the behavior we consider a toy  model. In this model the  signal from a single cap (i.e., a single accelerating spherical cap) is described by the function $h(t)$. The mass ejection function, $\dot m(t)$, describes the rate of ejection of shells. We choose a simple non-trivial model which involves two timescales:
\begin{equation}
\label{eq:toy}
    h(t) =  \begin{cases}  0  \quad & t < 0 \ , \\ 
    [{t}/{\tilde t_{\rm o} (\theta_{\rm v},\theta_{\rm j})} ]  h_0 (\theta_{\rm v}, \theta_{\rm j}) \quad & 0 \leq  t \leq \tilde t_{\rm o}(\theta_{\rm v},\theta_{\rm j}) \ , \\  
    h_0(\theta_{\rm v}, \theta_{\rm j}) \quad & t >\tilde t_{\rm o}(\theta_{\rm v},\theta_{\rm j}) \ ;\end{cases}
\end{equation}
and
\begin{equation}
\label{eq:toya}
\dot m (t) =  \begin{cases}  0 \quad & t < 0 \ , 
    \\ \dot m_0 , \quad & 0 \leq t \leq t_{\rm inj}  \ ,
    \\ 0 \quad &  t > t_{\rm inj}  \ . \end{cases}
\end{equation}
where $\tilde t_{\rm o}(\theta_{\rm v},\theta_{\rm j}) \equiv (1-\beta \cos(\theta_{\rm v}+\theta_{\rm j})) t_{\rm acc}$ and $h_0 (\theta_{\rm v},\theta_{\rm j})$ is the (ZFL) jump of the GW amplitude, given by Eq. \ref{eq:h_int}.} The combined GW signal is given by the convolution of the two functions. 
The amplitude of the Fourier transform at the crossover frequency is $h_0(\theta_{\rm v},\theta_{\rm j}) / {f_{\rm c} }$. The observed crossover frequency is now determined by two timescales, and one of those, $\tilde t_{\rm o}(\theta_{\rm v},\theta_{\rm j})$, varies with $\theta_{\rm v}$. 
Fig. \ref{fig:tj_and_t0} depicts the crossover diagrams determined by the simple model of eqs. \ref{eq:toy} - \ref{eq:toya} .

For $t_{\rm inj}  \ll \tilde t_{\rm o}$, we recover the previously described crossover diagram. For $t_{\rm inj}  > \tilde t_{\rm o} $, the injection time  acts as an upper cutoff on the crossover frequency. For $t_{\rm inj}  \gg t_{\rm acc}$ (which implies $t_{\rm inj}\gg \tilde t_{\rm o}$ for all observers) the crossover diagram is reduced to a single frequency determined by $t_{\rm inj} $, independent of $\theta_{\rm v}$. 
Clearly, if several timescales are involved in the function $\dot m(t)$, it is the longest one that determines the crossover frequency. The shorter timescales will only affect the higher-frequency range of the Fourier spectrum.
 
\section{An Example - GWs from GRB light curves}
\label{sec:GRBs}
The results of the previous section were based on a simplified model for the mass flux of the jet $\dot m(t)$. Here, we examine a possibly more realistic description. 
For this we consider GW emission from GRB jets assuming that the GRB light curves follow  $\dot m(t)$ to some extent.

Specifically, Kobayashi et al., \cite{kobayashi} have shown that within the internal shocks model \cite{NPP92,RM1994,SariPiran97} the GRB light curve is related to $\dot m(t)$. This relation is not one-to-one and, moreover, current understanding suggests that the temporal structure may originate in the interaction of the jet with stellar material (in long GRBs) or with the ejecta (in short ones).   Still, in the following,  we use the GRB light curves as indicators for  $\dot m(t)$, and estimate the corresponding GW signal. For a given acceleration model, the Fourier transform of the GW signal will be proportional to the convolution of the Fourier transform of the GRB light curve with the GW signal of a single shell $h_{\rm cap}(t)$.
We calculate, under these assumptions, the average GW spectra for long and short GRBs observed by the Burst and Transient Source Experiment (BATSE). We use a Fourier transform,  $\tilde h(f)$, of a single accelerating spherical cap (Eq. \ref{eq:crossover}), with $\tilde f_{\rm o}  \equiv {1}/{\tilde t_{\rm o}(\theta_{\rm v},\theta_{\rm j})}$ and $g(f) = f^{-\alpha}$ with $\alpha> 3/2$ being the high-frequency power law from the acceleration model Fourier transform. The following calculations will proceed with a general $\alpha$, which is determined by the specific acceleration model.

\subsection{Long GRBs} 
Beloborodov et al. \cite{Beloborodov} calculated the average Fourier transform, $C_{\it l}(f)$,  of 527 long GRB light curves observed by BATSE:
\begin{equation}
    \label{eq:grb}
    C_{\it l} (f) \propto  \begin{cases} {\mathrm const.} , \quad \quad f < f_{\it{l}}  \\ f^{-0.75} ,\ \ \quad f > f_{\it{l}}   \end{cases}\ ,
\end{equation}
where the spectrum changes its slope at $f_{\it{l}}  \approx 0.01 Hz$. 
 As such, the Fourier transform of the  GW signal, $\tilde h_{\it l} (f)$, will be  (see Fig. \ref{fig:h_short}):
\begin{equation}
    \label{eq:grb_conv_long}
    \tilde h_{\it l} (f) = C_{\it l} (f) \tilde h(f)  \propto  \begin{cases} f^{-1} ,\quad \quad \quad f < f_{\it{l}}  \\ f^{-1.75} ,\quad \quad f_{\it{l}}  < f < \tilde f_{\rm o}  \\ f^{-0.75-\alpha} , \quad \tilde f_{\rm o}  < f \ . \end{cases}
\end{equation}
The low-frequency behavior of the Fourier transform always behaves like  $1 / f$. The introduction of a new timescale means that there are  two crossover frequencies, between three different power laws. In the intermediate range $f_{\it{l}}  < f < \tilde f_{\rm o} $ the power law is determined purely by the GRB light curve, namely by the mass injection function. The unknonw high-frequency power law of the acceleration model, $\alpha$,  appears only at frequencies higher than the acceleration model's crossover frequency.

\subsection{Short GRBs}
\begin{figure}
\includegraphics[width=0.45\textwidth]{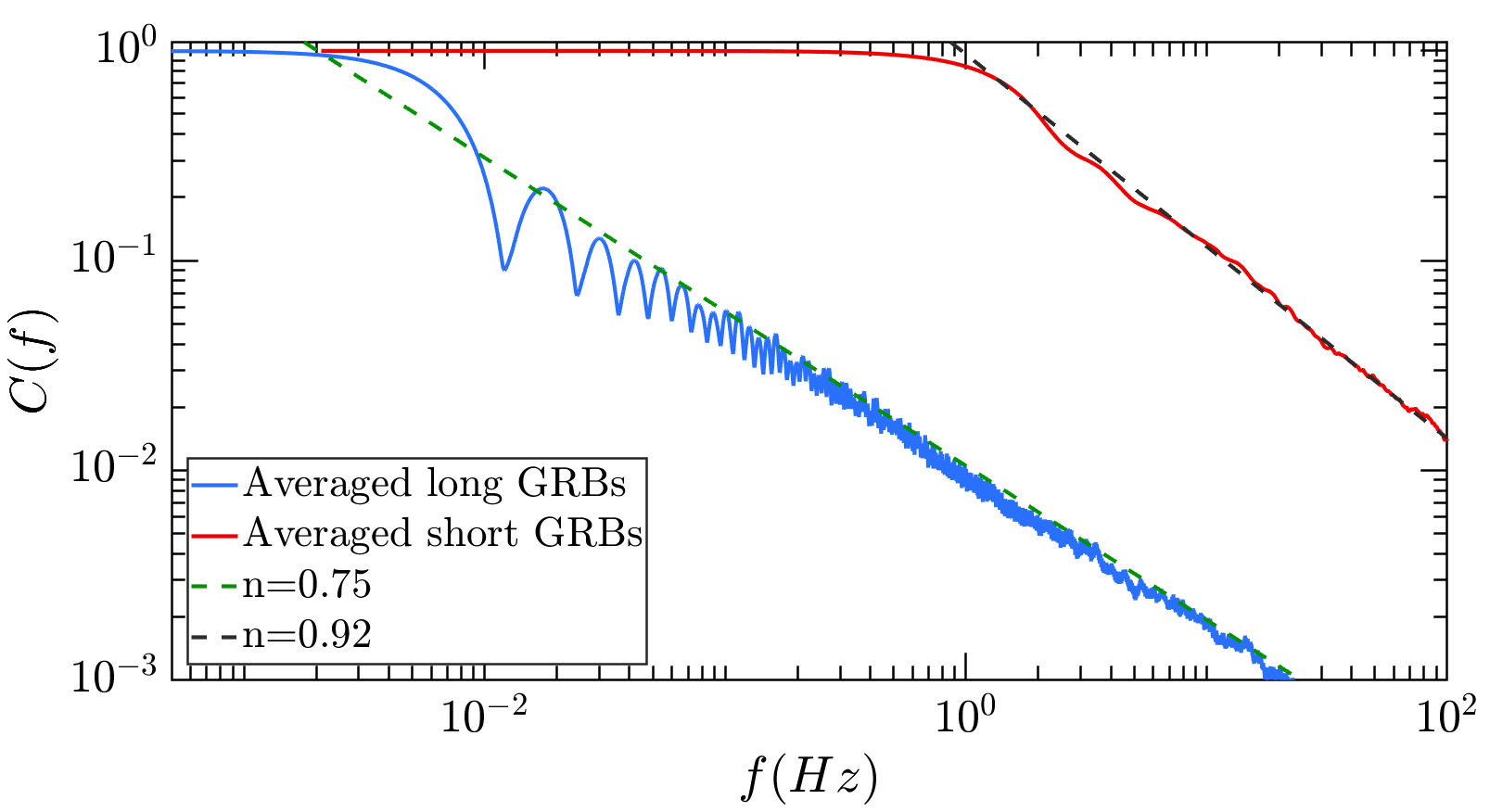}
\caption{The averaged Fourier transform of BATSE's long GRBs, vs. that of BATSE's TTE short GRB catalogue. Power law fits for are shown in dashed lines. }
\label{fig:tte}
\end{figure}

{
The temporal behavior of short GRBs is different from that of long ones. We repeated the above procedure, now using the TTE dataset from BATSE's measurements, which details the arrival times of individual photons. Using a bin size of $10 $msec, finding the average Fourier transform of  short GRBs:  
\begin{equation}
    \label{eq:grb}
    C_{\it s} (f) \propto  \begin{cases} {\mathrm const.}  , \quad \quad f < f_{\it{s}}  \\ f^{-0.92} ,\ \ \quad f > f_{\it{s}}  \end{cases}
\end{equation}
The high frequency power law of the short GRBs power spectrum is stiffer, and their break frequency is higher, at $f_{\it{s}}  \approx 1 Hz$, corresponding to the timescale of an average short GRB (see Fig. [\ref{fig:tte}]). The Fourier transform of the corresponding GW signal of a short GRB  takes the form:
\begin{equation}
    \label{eq:grb_conv_short}
    \tilde h_{\it s} (f) = C_{\it s} (f) \tilde h(f)  \propto  \begin{cases} f^{-1} , \quad \quad \quad f < f_{\it{s}}  \\ f^{-1.92} ,\quad \quad f_{\it{s}}  < f < \tilde f_{\rm o}  \\ f^{-0.92-\alpha} , \quad \tilde f_{\rm o}  < f \end{cases}
\end{equation}
This result holds only if $f_{\it{s}} < f_{obs}$. 
The Fourier transform of the short GRBs, $C_{\it s}$,  allows for the interesting scenario in which  
this is not the case, and 
the observed acceleration timescale $\tilde t_{\rm o}(\theta_{\rm v},\theta_{\rm j})$ may be longer than the mass ejection timescale $t_{\rm inj} $. In this case, the form of the GW's Fourier transform will be slightly different. At low and very high frequencies, the Fourier transform  still behaves like  $1/f$  and  $f^{-0.92-\alpha}$, correspondingly. However, in the intermediate frequency range  $\tilde f_{\rm o}  < f < f_{\it{s}} $, the power law will be different:
\begin{equation}
    \label{eq:grb_conv_short_reversed}
    \tilde h_{\it s} (f) = C_{\it s} (f) \tilde h(f)  \propto  \begin{cases} f^{-1} ,\quad \quad \quad f < \tilde f_{\rm o}  \\ f^{-\alpha} ,\quad \quad \quad \tilde f_{\rm o}  < f < f_{\it{s}}  \\ f^{-0.92-\alpha} , \quad f_{j} < f \end{cases}
\end{equation}
The two cases are illustrated in Fig. \ref{fig:h_short}, where we plot the Fourier transforms of two short GRB's GWs using the Fireball acceleration model: one with $t_{\rm inj}  > \tilde t_{\rm o}(\theta_{\rm v}, \theta_{\rm j})$, and one with $t_{\rm inj}  < \tilde t_{\rm o}(\theta_{\rm v} , \theta_{\rm j})$. As it turns out, for $\alpha \approx 2$ the power laws of the intermediate frequency range in both cases are quite similar.
}

% \del{
% \begin{figure}
% \includegraphics[width=0.45\textwidth]{Fourier_N_steps.png}
% \caption{\del{The function $h(t)$ with $t_{\rm acc}=1$ is convolved with two repeating rectangular functions: one with $N=1$, and one with $N=5$. Both have $t_{\rm inj} =100$. The spectral density is plotted for both convolutions. The crossover frequency, and the behavior in frequencies lower than the crossover region, remains unchanged. For $N=5$ , the behavior in higher frequencies is changed, including a new peak corresponding to the timescale of a single rectangle, $\frac{t_{\rm inj} }{N}$. }}
% \label{fig:Fourier_N_steps}
% \end{figure}
% % }
% \del{
% \begin{figure}
% \includegraphics[width=0.45\textwidth]{Fourier_N_steps_norm.png}
% \caption{\del{The Fourier transform of Fig. \ref{fig:Fourier_rect}), multiplied by $f$. For $f<f_{\rm c} $, the transform behaves as $\propto 1/f$, and thus $\tilde h(f) \cdot f$ is a constant. Both $N=1$ and $N=5$ break at the same frequency, which corresponds to $t_{\rm inj} $.}}
% \label{fig:Fourier_N_steps_norm}
% \end{figure}
% }

\begin{figure}
\includegraphics[width=0.45\textwidth]{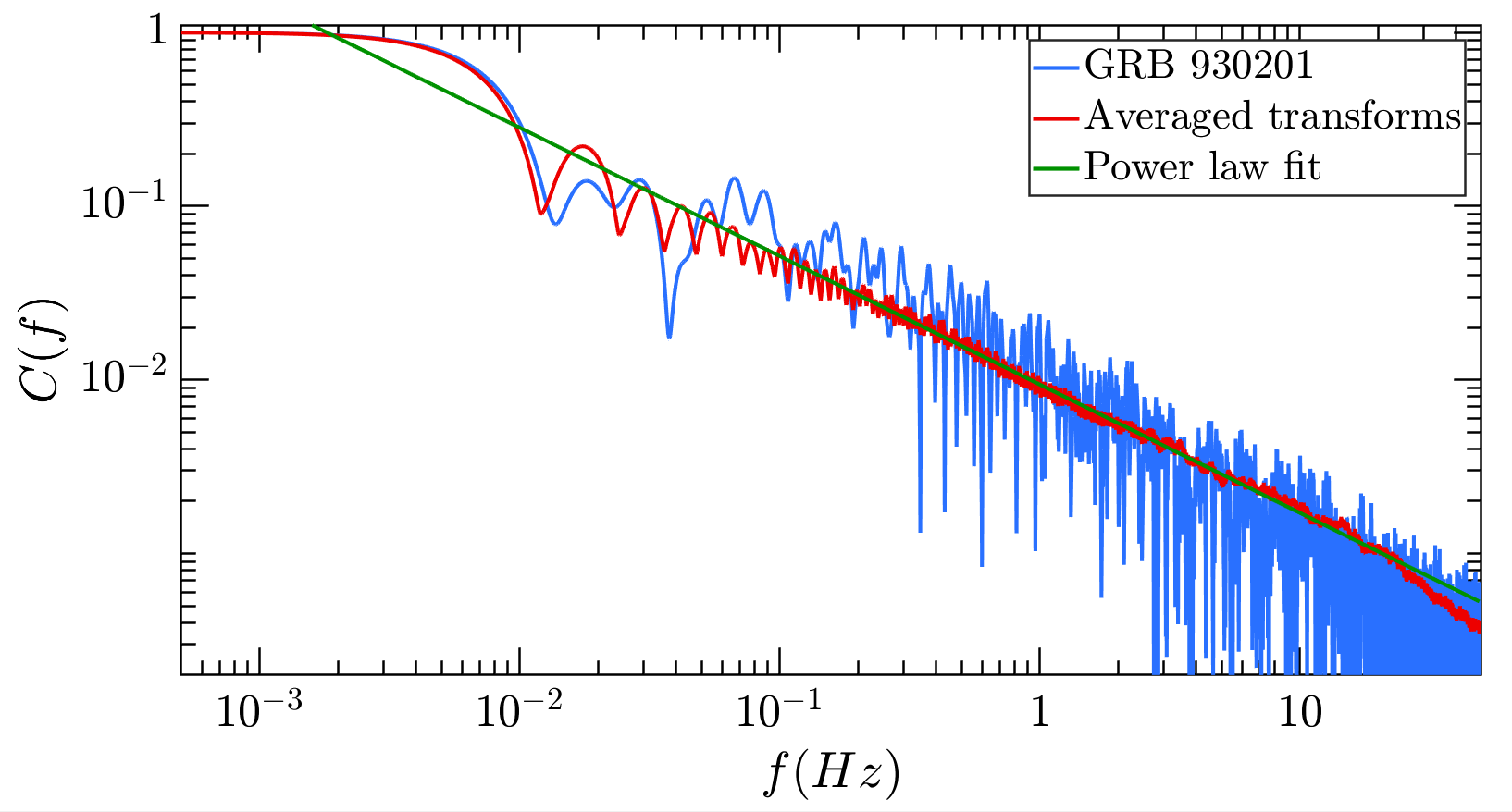}
\caption{ The Fourier transform of the light curve GRB 930201, one of the brightest bursts observed by BATSE from the BATSE data (blue).   The averaged Fourier transforms of all bursts observed by BATSE (red).  A power-law fit $f^{-n}$ for the average of the Fourier transforms, with $n = 0.75$ (green). When averaging over many different bursts, the noise components cancel out. 
The frequency where the the Fourier transofrm of GRB 930201 levels out to a constant is determined by the duration of the GRB.}
\label{fig:BATSE_2156}
\end{figure}

\begin{figure}
\includegraphics[width=0.45\textwidth]{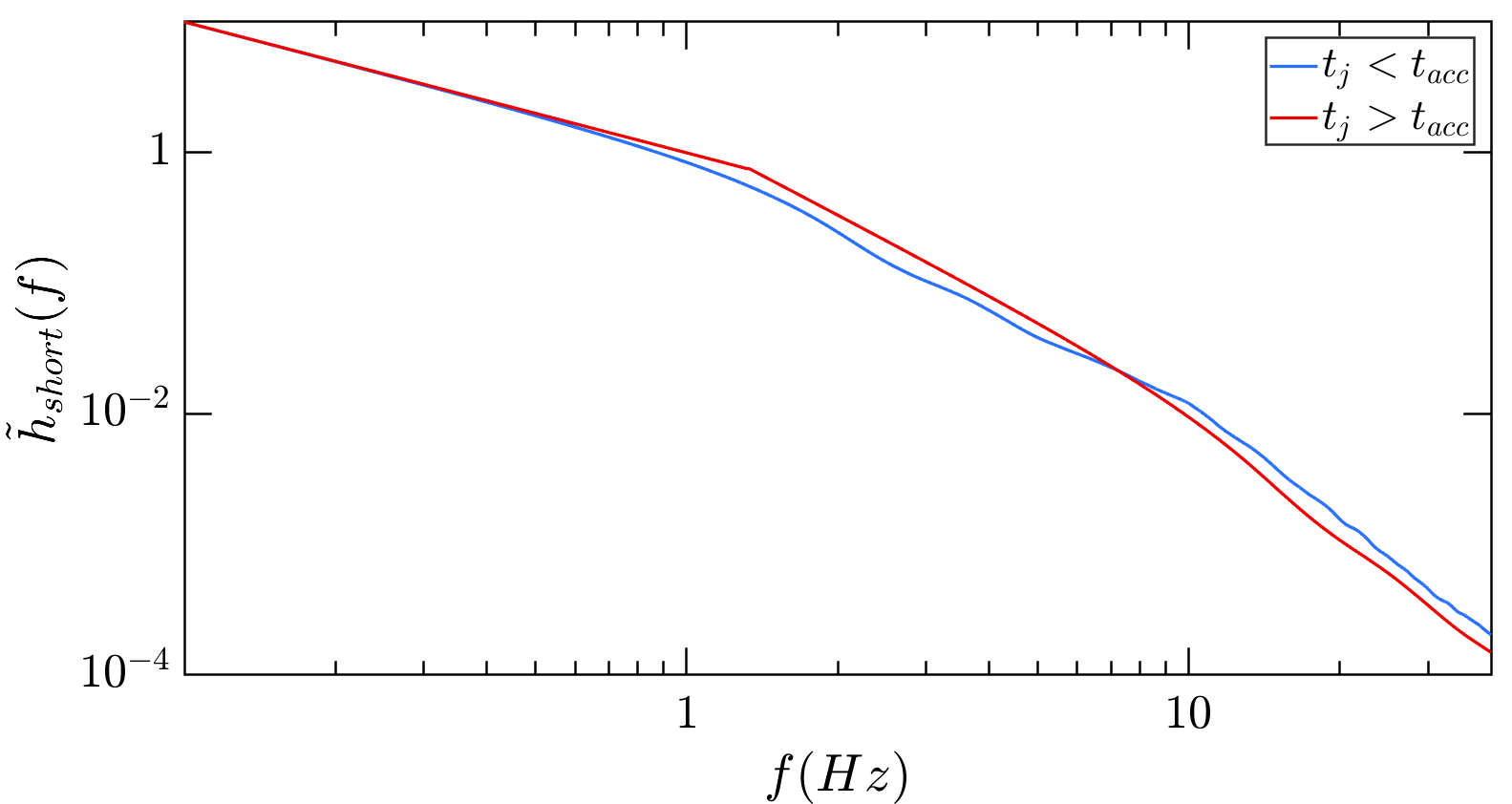}
\caption{The Fourier transform for a short GRB's GW  calculated  with $t_{\rm inj}  =1 {\rm sec} , \tilde t_{\rm o}(\theta_{\rm v})=0.1 {\rm sec}$ (blue), and  with $t_{\rm inj}  = 0.1 {\rm sec} ,\tilde t_{\rm o}(\theta_{\rm v}) = 1 {\rm sec}$ (red). The power laws in the intermediate frequency region between $f_{\it{s}} $ and $\tilde f_{\rm o} $ are slightly different for the two cases: $\tilde h \propto f^{-1.92}$ for $t_{\rm inj}  > \tilde t_{\rm o}(\theta_{\rm v})$, and $\tilde h \propto f^{-\alpha}$ for $t_{\rm inj}  < \tilde t_{\rm o}(\theta_{\rm v})$.}
\label{fig:h_short}
\end{figure}

\section{Detectability}
\label{sec:detectability}
 
When estimating the detectability of a GW signal, we have to compare the expected $S(f)$ to the detector's sensitivity curve,  $S_{\rm det}$, taking into account both the amplitude and the relevant frequency range. As we have seen in \S \ref{sec:temporal}, for jet GW signals S(f) is always a decreasing function of the frequency.  At the lowest frequency range $S(f) \propto f^{-1/2}$, while at higher frequencies (above the relevant crossover frequency) it decrease faster. Hence, a typical  low-frequency detector will be most sensitive to a jet GW  signal at  its  lowest end of its frequency response. For our purposes we can define this point as the lowest frequency below which $S_{\rm det}$ is steeper than $f^{-1/2}$.  A similar condition holds for a high-frequency detector (that is, above a crossover frequency) for which we replace the power $f^{-1/2}$ by the corresponding frequency dependence of the spectral density. 

Not surprisingly, like almost any relativistic GW source, the maximal amplitude of the  jet GW is of order 
\begin{equation}
h \approx \frac{ G \cE }{ c^4 r } = 3 \times 10^{-25} \bigg(\frac{\cE}{10^{51} ~{\rm erg}}\bigg)~\bigg(\frac{100 {\rm Mpc}}{r} \bigg) \ . \label{eq:magnitude}
\end{equation}
%where we have used somewhat optimistic values for the energy and the distance.
For a one-sided jet this estimate is valid for an observer that is at optimal angle, namely at $\theta_{\rm v} \approx \theta_{\rm j}$. For two-sided jets, this estimate is valid for most  observers apart from those along the jets ( $\theta_{\rm v} <  \theta_{\rm j}$).  
Different observers will, however, observer different characteristic frequencies as discussed earlier, with the relevant frequency is the lowest crossover frequency, $f_{\rm c}$. 

\subsection{GW from GRB jets}
\label{sec:GRBs.detect}

GRB jets are the most natural sources for these kind of GWs. 
For an optimal observer near the jet, when considering the estimates based on the GRB light curves discussed in \S \ref{sec:GRBs}, the crossover frequency is dominated by $t_{\rm inj}$ for both long and short GRBs. Thus, $f_{\rm c}=f_{\it{l}}= 0.01$ Hz  for the long and $f_{\rm c}=f_{\it{s}}=1$ Hz for short GRBs.  This frequency range puts the events below the frequency limits of current LIGO-Virgo-Kagra,  but around the capability of the planned BBO \cite{BBO} and DECIGO \cite{Decigo}. Observers further away from the jet axis will see lower characteristic frequencies, which are even more difficult to detect. 

%More optimistic GRB estimates in the conclusions. 

%A low frequency detector (with a typical frequency range below $f_{\rm c} $) will detect such event if for some frequency in its range, $f$, it satisfies $S_{det} (f )< S(f_{\rm c} ) (f_{\rm c} /f)^{1/2}$.    The detector will measure a step function. Higher frequency detectors that are sensitive in the range above $f_{\rm c} $ can detect the relevant and interesting temporal structure that exists beyond a simple step function. 
As seen in the crossover diagram (Fig. \ref{fig:crossover_pp}), any potential increase in the frequency of the spectral density due to the boost of the crossover frequencies for observers close to the jet's axis will be more than balanced out by the anti-beaming of the GW amplitude, such that the spectral density never benefits from observation-angle effects. 
Short GRBs have higher crossover frequencies and hence are somewhat easier to detect. These bursts are observed from typically nearer distances since they are intrinsically weaker and hence their observed rate is lower.   However  their intrinsic rate is larger by about a factor of ten than the rate of long ones. Still, due to current LIGO-Virgo-Kagra lower frequency threshold in the $10$s Hz range, which is above the expected crossover frequencies of short GRBs and definitely long GRBs, it is unlikely for any GW signal from either short  or long GRB jet to be detected by these detectors. While these GRB jets GW signals are within the frequency range of BBO and DECIGO, most GRBs  take place at distances that are  beyond the detection horizon. 

\subsection{GW 170817A}
\label{sec:170817}
\begin{figure}
\includegraphics[width=0.45\textwidth]{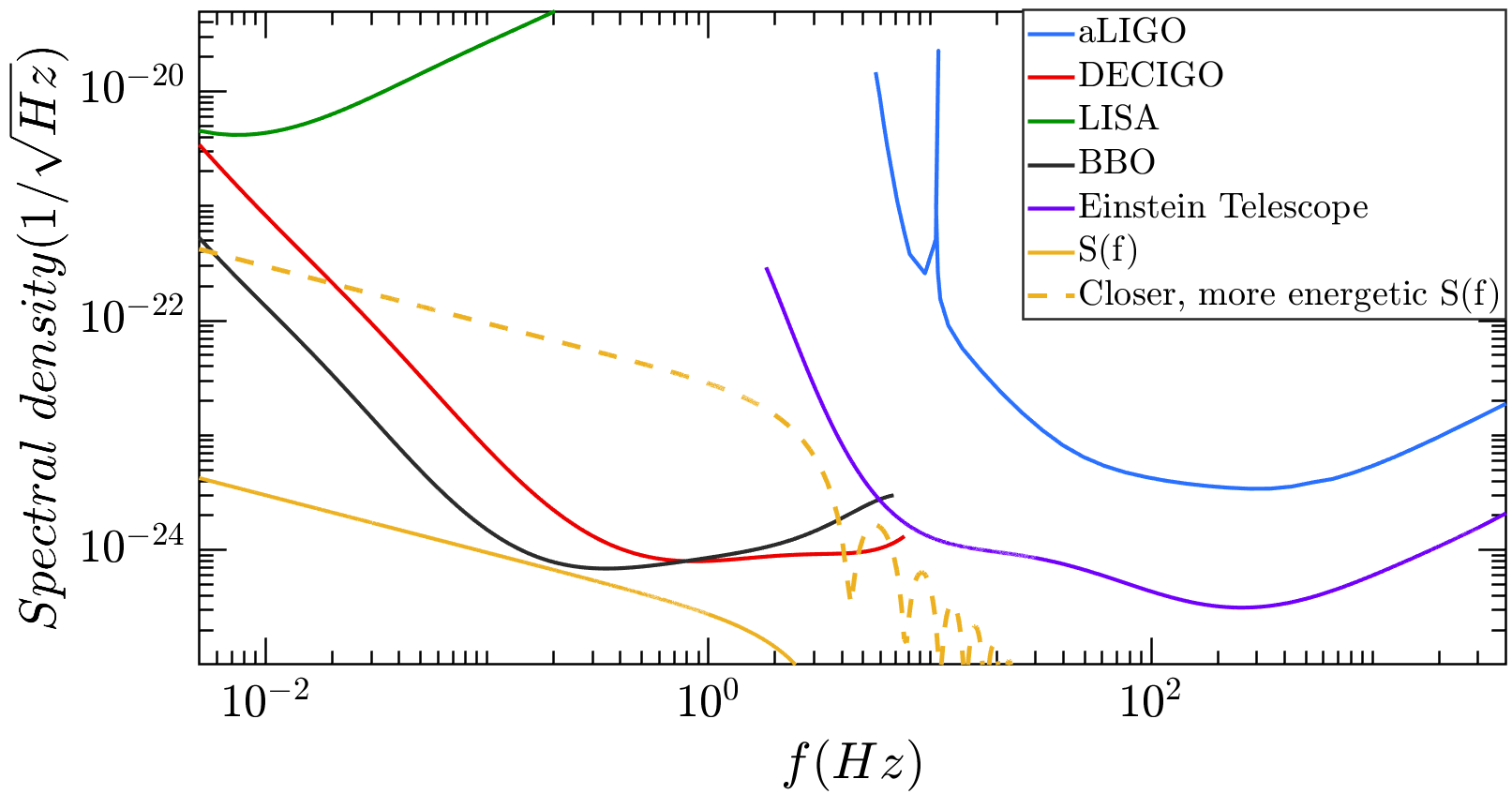}
\caption{The calculated spectral density for our fiducial model for GW170817, $S(f)$, compared with the sensitivity thresholds of GW detectors taken from http://gwplotter.com. {The dashed line shows the GW emission from the same source, only 10 times closer and 10 times more energetic. Such a signal would correspond to a CCSNe jet }}
\label{fig:LIGO}
\end{figure}

At $\approx 40$ Mpc, GW170817A was an exceptionally nearby binary neutron star merger. The merger GW signal was  accompanied by a short (albeit atypical -- see e.g. \cite{Kasliwal17,Kasliwal17,Nakar20}) GRB. The event and its afterglow signature were extremely well observed, and we have good estimates for most of its parameters. The jet properties are  $\cE  \approx 10^{50} $erg,  $\theta_{\rm v}  \approx 20^o$, $\theta_{\rm j}\approx 5^o$.
Other parameters, and in particular $t_{\rm acc} $ and $t_{\rm inj}$ that are most relevant for our analysis, are less known. The injection duration $t_{\rm inj}$, is capped from above by the duration of the observed $\gamma$-rays. However, as those arose from a cocoon shock breakout \cite{Kasliwal17,Gottlieb18} the observed duration gives only an upper limit on $t_{\rm inj}$ .  In the following we assume that $t_{\rm acc}< t_{\rm inj} =  1$ sec. 
$\Gamma$, is also unknown but  it only factors into the result through $\cE$, since $\Gamma^{-1} \ll \theta_{\rm j}$: hence, it is unimportant.  Given the viewing angle and the jet angle, it was also ideally positioned in terms of the strength of the GW signal from its jet. That is, we were not within the anti-beamed jet's cone but not too far from it either. 
Still, the jet GW that we consider here could not have been detected by current detectors. 
Fig. \ref{fig:LIGO}, depicts the spectral density of GW170817 compared with the sensitivity thresholds of GW detectors \cite{Moore_2014}. We find that the GW would have been  detectable by the Big Bang Observer (BBO)\cite{Crowder_2005}, and would have been marginally detectable by DECIGO\cite{Sato:2017dkf} as we discuss below.. 

We quantify detection distances by considering the signal-to-noise ratio, $\rho$, of a certain GW signal, with Fourier transform $\tilde h(f)$ \cite{Moore_2014}:
\begin{equation}
    \rho^2 = 4 \int_{-\infty} ^\infty \frac{\tilde h(f)^2}{S_n (f) ^2} df ,
    \label{eq:snr}
\end{equation}
where $S_n (f)$ is the detector's noise amplitude.

%\begin{figure}
%\includegraphics[width=0.45\textwidth]{snr_vs_distance.png}
%\caption{The signal-to-noise ratio of a GW170817-like signal, emitted by a source at a distance $r$, as measured by different detectors. }
%\label{fig:snr_vs_distance}
%\end{figure}
%In Fig. \ref{fig:snr_vs_distance}, we calculate the SNR of a GW signal with all of GW170817's aforementioned parameters, except that we allow the distance of the source, $r$, to vary. We define the maximal detection distance $r_d$ for a certain detector as the distance at which $\rho=1$. 

We find that the most suitable detector for observing jet GWs is BBO, with a detection horizon of $r_d = 75 $Mpc. DECIGO closely follows, with $r_d = 40 $Mpc. The Einstein telescope has $r_d= 600 $kpc, and LISA is at $r_d = 80$ kpc. 
Ultimate DECIGO which will be about hunderd time more sensitve than DECIGO will detect such events from distances of a few Gpc, that is up to  $z=0.5$. 
These distances scale linearly with the jet's energy: a jet with a short duration like  GW170817 but   with $E=10^{51}$ erg will be detectable by DECIGO up to a distance of $  400$ Mpc, etc.

%\begin{figure}
%\includegraphics[width=0.45\textwidth]{rd_vs_fc.png}
%\caption{DECIGO's maximal detection distance for a GW170817-like signal, for different crossover frequencies. Above a certain crossover frequency, the maximal detection distance doesn't increase.}
%\label{fig:rd_vs_fc}
%\end{figure}

%Generally, we find (see Fig. \ref{fig:rd_vs_fc}) that a
A higher  GW crossover frequency, $f_{\rm c} $, increases the maximal detection distance $r_d$. Notably, however, $r_d$ approaches an asymptotic value, and increasing $f_{\rm c} $ above a certain detector-specific threshold does not change that detector's maximal detection distance. This is because the integral in Eq. \ref{eq:snr} is dominated by the part of the GW's Fourier transform which is within the detector's frequency band. If $f_{\rm c} $ is higher than this band, then the integral is dominated by the low-frequency $\sim 1 / f$ behavior of the transform, which is independent of $f_{\rm c} $. When $f_{\rm c} $ is within the detector's frequency band, the SNR will be reduced, due to the integration over the higher-frequency region of the GW, which behaves as $\sim 1 / f^2$.
%This last result is important when one considers the important problem of deducing the temporal structure of the jet from the GW observation. 
%However, to learn about the temporal structure, the crossover frequency would have to fall within the detector's frequency band. Fig. \ref{fig:rd_vs_fc} should be interpreted as the price in maximal detection distance one has to pay, in order to observe GWs with lower crossover frequencies.

\subsection{Jets in Core Collapse SNe and low-luminosity GRBs}
\label{sec:CCNe}

The prospects for CCSNe-related GW detection are much more optimistic. Shortly after the discovery of the first low-luminosity GRB 980415 (that was associated with SN98bw) it was suggested \cite{Kulkarni98,MacFadyen01,Tan01} 
that the emission arose from shock breakout following an energetic jet that was choked deep in the accompanying star. Later on it was realized  that, while the detection rate of low-luminosity GRBs is much lower than that of regular long GRBs, their actual rate is orders of magnitude larger \cite{Soderberg06,Bromberg11}. The detection rate is small because, given their low luminosity, they are detected only from relatively short distances. 
More recently,  Piran et al. \cite{Piran17} have shown that a significant fraction of CCSNe (that are not associated with GRBs)  contain an energetic ($\sim 10^{51}$ erg) choked relativistic jet. 
While this jet is relativistic, it is chocked inside the star depositing its energy into a cocoon. Upon breakout the cocoon material is observed as a  high velocity (0.1-0.2c) material that engulfs the supernova and can be detected within the first few days. Such signatures have been detected as early as 1997 \cite{Mazzali00} in SN 1997EF and in several other SNe since then. This suggestion was 
nicely confirmed with the exquisite observations of this high velocity material in  SN 2017iuk by \cite{Izzo19,Nakar19}.  
If such relativistic jets are associated with a significant fraction of  CCSNe then, as the supernova rate is significantly larger than GRB rate \cite{Piran_fireball}, we can expect  much nearer jets that would be sources of such GWs. 

Comparing relativistic SNe Jets with GRB jets, we estimate $h$ to be a factor of 100-1000 larger than the one estimated for short GRBs: a factor of 10 in the distance (tens of Mpc vs. hundreds of Mpc) and a factor of 10-10 in energy  ($10^{51}$ erg vs.  $10^{49-50}$ erg). 
Thus, we expect amplitudes of $3 \times 10^{-24}$ (see Eq. \ref{eq:magnitude}). Unfortunately, for these events we don't have a good clue on $t_{\rm inj}$. A best guess is that it will be of the same order as the one estimated in long GRB, namely of order of a few tens of seconds. Thus, the corresponding crossover frequency would be around 0.01 Hz. However, on average we will observe these events from a large viewing angle, and in this case the crossover frequency would be even lower. The exact value will depend on $t_{\rm acc}$, and in turn on the unknown nature of the acceleration process.  

%For the prospects of joint EM-GW detection, the form of the crossover diagram is luckily not so limiting, thanks to the anti-beaming behavior of jets with a finite opening angle. Going again by GW170817's example, and again assuming $t_{\rm acc}  \ll t_{\rm inj} $, the scaling of the spectral density with $\theta_{\rm v}$ is determined entirely by Eq. \ref{eq:h_int}. One can see that, for $\theta_{\rm j}=5^0$, the GW amplitude is maximized at $\theta_{\rm v}=21^0$, which almost coincides with the actual observation angle of the event. The amplitude will only go down to $50\%$ of maximum for observers within the cone $\theta_{\rm v} = 5^0$. In this sense, even though the GW emission is anti-beamed and the EM emission is beamed, the reduction in GW amplitude in the vicinity of $\theta_{\rm v} \geq \theta_{\rm j}$ is marginal.

%If the assumption $t_{\rm acc}  \ll t_{\rm inj} $ does not apply, the situation is slightly worse. Depending on the ratio $t_{\rm inj}  / t_{\rm acc} $. In the extreme case where $t_{\rm inj}  \ll t_{\rm acc}  $, the spectral density drops to $50\%$ of maximum for observers within $\theta_{\rm v}=20^0$ (see Fig. \ref{fig:crossover_pp}) and the related discussion).

\subsection{Contribution to the GW background}
\label{sec:GWbackground}

{  The relativistic jets that arise from GRBs (both long and short) and hidden jets in SNe produce a continuous background of jet-GW waves at frequency range of $\sim  0.01-1  $Hz depending on the specific source.   Both long and short GRBs are rare and won't make a significant contribution to such a background. However,  SNe take place at a rate of about one per second in the observable Universe. If a significant fraction of SNe harbor energetic jets the time between two such cosmological events, a few seconds,  will be comparable to the characteristic time scale of the GW signals from these jets (assuming that the hidden jets in SNe are similar in nature to GRB jets).   Depending on the ratio of the time between events and the characteristic frequency of the jet-GW  signal we expect either a continuous background, as expected from the GW background from merging binary neutron stars,   or a pop-corn like signature, as expected for the GW background from merging binary black holes \cite{LIGOstochastic}.   With a typical cosmological distance  of a few Gpc the corresponding  amplitude of this jet-GW background is $h \approx 10^{-26} \cE/(10^{51} {\rm erg})$. }

\section{Discussion}
\label{sec:discussion}
We have obtained the qualitative and quantitative behavior of the amplitude, the angular distribution of both $h$ and $dE_{\rm GW} /d\Omega$, and the Fourier transform of the GW signal of an accelerated jet with 
an opening angle $\theta_{\rm j}$. The signal is anti-beamed away from the direction of the jet. The anti-beaming angle is $\max(\Gamma^{-1},\theta_{\rm j}$).
Like typical relativistic GW sources, the amplitude is of order $G\cE/c^4 r $. However, unlike other sources, the signal here is of a memory type, rising to this amplitude on a characteristic time scale. 
The signal can be approximated as a step function when considering detectors whose typical response frequency is much lower than the characteristic crossover frequency of the jet.  This last feature is of course problematic, as it might be difficult to distinguish this signal from other step functions that may arise in GW detectors. We won't explore the experimental/observational aspects of this question. 

The light curve depends on two  timescales: the acceleration timescale $t_{\rm acc}$, and the mass ejection time $t_{\rm inj} $. 
The spectral density $S(f)$ is monotonically decreasing with the frequency. It is broken into at least two power laws: the lower frequency region is proportional to $f^{-1/2}$, and the higher frequency region is proportional to $f^{-1/2-\alpha}$, with $\alpha>3 / 2$. The spectral density is characterized by the crossover region, $f_{\rm c}$, which corresponds to the longest relevant timescale.  Since $S(f)$  decreases monotonically with frequency, the crossover region is  a good indicator as to whether a given GW's signal can be measured by a specific detector.

The universal form of the 'crossover diagrams'  describe how the frequency and the amplitude of the spectral density shift due to the dependence of the observed amplitude and frequency on $\theta_{\rm v}$. We calculated these 'crossover diagrams' for  a point particle, a jet with a finite opening angle,  a double-headed jet, as well as for jets with both $t_{\rm acc}$ and $t_{\rm inj} $. For $t_{\rm inj}  \gg t_{\rm acc}$, the crossover diagram is reduced to a single characteristic frequency for observers at all angles. 

Assuming that the observed GRB light curves are proportional to the jet's mass ejection function $\dot m(t)$ and assuming  a  specific acceleration model, we  calculated possible examples of expected  GW signals from long and short GRBs jets. As expected, we find that the composite Fourier transforms are monotonically decreasing, and that they are described by two crossover frequencies, between three power laws. One crossover frequency is associated with $t_{\rm inj} $, and the other is  associated with $t_{\rm acc}$. It is important to note, however, that these estimates should be considered just as examples. 

Recent understanding of jet propagation in dense media 
%[e.g. \cite{Gottlieb21}] suggests that it is possible and even likely that propagation effects of the jets within the surrounding matter (stellar envelope for Collapsars and ejecta for mergers) rather than processes taking place at the central engine dictate the fluctuations observed  in GRBs light curve.  
suggests that the injection  must be longer than the observed duration of the GRB \cite{Bromberg11}. 
Thus, the latter puts a lower limit on $t_{\rm inj}$.
%In these case the latter provide only upper limits on $t_{\rm acc}$ and $t_{\rm inj} $. If those are much shorter than the GW jet signal might be even  within 
%LIGO-Virgo-Kagra frequency range. 
However, the light curves of short GRBs suggest that in many cases mergers produce jet that are choked inside the merger ejecta. Those events are not accompanied by a short GRB \cite{MoharanaTP}. In such a case, $t_{\rm inj} $ can be much shorter (this is the reason that the jet was choked), and the corresponding GW signal will have a higher frequency. 
%If $\dot m(t)$ involves several timescales, then its longest timescale determines the lowest crossover frequency. The shorter timescales will only affect the higher-frequency behavior, starting with the frequency associated with them.

As an example, we calculated the gravitational waveform of the GW emitted by the jet associated with GW170817 under the previous assumptions. Using the event's parameters, we found that the jet's GW could have been observed by BBO and DECIGO. Within the limiting assumptions that the duration of the burst and the observed $\gamma$-ray light curve reflect the injection time, the relevant frequencies are quite low, and indeed BBO and DEGIGO are the most suitable detectors for observing GWs from similar short GRB jets. Anti-beaming will, however, make it unlikely that we would observe both the $\gamma$-rays and the GWs. However, other multimessenger signals, and in particular GWs from the merger itself, would accompany such an event triggering our attention and providing a additional significance to the detected GW signal. It is interesting to remark that the jet launching can be delayed by as much as a second after the merger, and, as such, this GW signal can be easily separated from the more ``regular" pre-merger GW emission, and even from the post-merger ringdown of the proto-neutron star and collapse to a black hole.  

While the detection prospects of a jet GW signature from short or long GRBs are not that promising, comparable or even more powerful relativistic jets also take place within some core collapse SNe.  The rate of these events is much larger, and correspondingly within a given observing  time frame they will take place at much nearer distances. Here the detection prospects are very promising once detectors in the sub-Hz are available. 
A detection would reveal features of jet acceleration in the vicinity of black holes that are impossible to find in any other way. 

\begin{acknowledgments}
We thank Ofek Birnholtz for providing us his code and for helpful comments and Ehud Nakar and Amos Ori for fruitful discussions. The research was supported by an advanced ERC grant TReX. 
\end{acknowledgments}

%\printbibliography
%\bibliography{research}
%\end{document}
\end{document}